\documentclass[a4paper,amsmath,amssymb,superscriptaddress]{article}
\usepackage[left=42mm,right=42mm]{geometry}
\usepackage[labelfont=bf]{caption}

\usepackage{eepic}
\usepackage{graphicx}
\usepackage{overpic}
\usepackage{latexsym}
\usepackage{bm}
\usepackage{amsmath}
\usepackage{amssymb}
\usepackage{amsthm}
\usepackage{microtype}
\usepackage[retainorgcmds]{IEEEtrantools}
\usepackage{caption}
\usepackage{subcaption}
\usepackage{array}
\usepackage{color}
\usepackage{txfonts}
\usepackage[T1]{fontenc}
\usepackage[utf8]{inputenc}
\usepackage{authblk}

\title{\Large MULTISTATE DYNAMICAL PROCESSES ON NETWORKS:
ANALYSIS THROUGH DEGREE-BASED APPROXIMATION FRAMEWORKS.}
\author[1]{PETER G. FENNELL\thanks{Corresponding author: pfennell@isi.edu}}
\author[2]{JAMES P.~GLEESON}
\affil[1]{\small Information Sciences Institute, University of Southern California, Marina del Rey, CA 90292, USA}
\affil[2]{MACSI, Department of Mathematics and Statistics, University of Limerick, Ireland}

\begin{document}

\maketitle

\begin{abstract}
  Multistate dynamical processes on networks, where nodes can occupy one of a multitude of discrete states, are gaining widespread use because of their ability to recreate realistic, complex behaviour that cannot be adequately captured by simpler binary-state models. In epidemiology, multistate models are employed to predict the evolution of real epidemics, while multistate models are used in the social sciences to study diverse opinions and complex phenomena such as segregation. In this paper, we introduce generalized approximation frameworks for the study and analysis of multistate dynamical processes on networks. These frameworks are degree-based, allowing for the analysis of the effect of network connectivity structures on  dynamical processes. We illustrate the utility of our approach with the analysis of two specific dynamical processes from the epidemiological and physical sciences. The approximation frameworks that we develop, along with open-source numerical solvers, provide a unifying framework and a valuable suite of tools for the interdisciplinary study of multistate dynamical processes on networks. 
\end{abstract}

\section{Introduction}
\label{sec:intro}
Networks are ubiquitous. From social networks to ecological networks, human contact networks to transportation networks and the World Wide Web, the existence of groups of units that are interconnected in some manner is common in our modern-day world~\cite{newman2010networks}. When modelling dynamical processes on networks~\cite{barrat2008dynamical,porter2016dynamical}, the most minimally complex models are those with binary state-spaces, i.e., models where, at each moment of time, a node can occupy either of two binary states~\cite{gleeson2013,liggett1985particle}. Nodes can change from one state to another, and do so probabilistically, with transition rates that depend on various local and global factors. Binary-state models have been applied to model many phenomena, with examples including the spread of computer viruses though the internet~\cite{kephart1993measuring}, global cascades in complex systems~\cite{watts2002simple} and the diffusion of opinions or sentiment through social networks~\cite{bass1969new}. 

However, many natural phenomena cannot be adequately modeled with a simple binary state-space and extensions of the modelling framework to multiple states are required. Examples abound in the field of epidemiology, where multistate models are used to model and predict the evolution of real epidemics such as the recent Ebola outbreak in West Africa~\cite{gomes2014assessing,lekone2006statistical}. Such models include various node states associated with the disease (such as susceptible, infectious, exposed) as well as other states associated to control and/or other actions (vaccinated, hospitalized, etc)~\cite{hethcote2000mathematics}, and form the basis for computational platforms used to predict and to develop control strategies for epidemics~\cite{balcan2010modeling}.

Despite the growing importance of multistate modelling, there exists no unifying theoretical framework to enable the analysis of how network structure affects the evolution of multistate dynamical processes. While specific frameworks have been developed in certain areas, such as for rumour spreading~\cite{de2016unifying} and the dynamics of interacting diseases~\cite{sanz2014dynamics}, a fully general approach that covers the whole class of mulstistate dynamical processes is still lacking. The purpose of this paper is to bridge that gap.

\begin{figure}
    \centering
    \includegraphics[width=0.9\columnwidth]{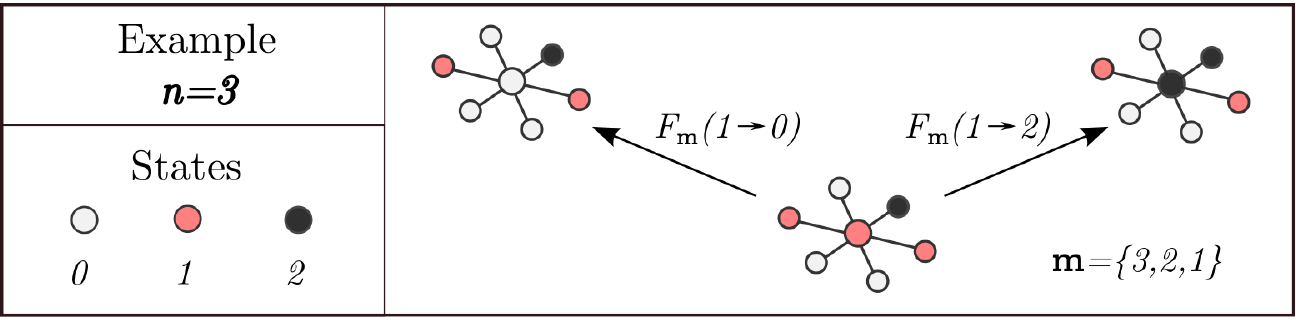}
    \caption{Schematic of a multistate dynamical process. In this example, nodes can be in one of three states (labelled 0, 1 or 2). Nodes change from their current state $i$ to a different state $j$ at a rate $F_{\mathbf{m}}(i\rightarrow j)$ that depends both on the state $i$ of the node and and the states of its neighbours, where the vector $\mathbf{m}=\{m_0,\dots,m_{n-1}\}$ encodes the number of neighbours in each of the $n$ states.}
    \label{fig:Multischematic}
\end{figure}

In this work, we consider the general class of continuous-time multistate dynamical processes (on undirected networks) that can be represented by rate functions $F_{\mathbf{m}}(i\rightarrow j)$, where $F_{\mathbf{m}}(i\rightarrow j)$ denotes the rate at which a node in state $i$ changes to state $j$ as a function of the number of its neighbours in the each of the $n$ multiple states of the dynamics (Fig.~\ref{fig:Multischematic}). This latter information is encoded in the vector $\mathbf{m} = \{m_0,\dots,m_{n-1}\}$, where $m_l$ is the number of neighbours of a node in state $l$. For a more compact representation, the rate functions can be combined into a rate function matrix $\mathbf{F}_{\mathbf{m}}$, where  $(\mathbf{F}_{\mathbf{m}})_{ij}=F_{\mathbf{m}}(i\rightarrow j)$. A very wide range of multistate dynamical process models from the literature can be represented in this form, and in Section~\ref{sec:multistate_dynamics} we give a brief review of such literature to illustrate the generality of our approach. 

Using this representation, we develop degree-based approximation frameworks for the analysis of multistate dynamics on uncorrelated networks, with a focus on the effects of the network degree distribution. We develop three such frameworks with varying levels of approximation, namely the mean-field, pair approximation and approximate master equation frameworks~\cite{gleeson2013,newman2010networks}. Furthermore, we develop software for the efficient numerical solution of these systems of differential equations, allowing for detailed analysis in the cases where the equations are not analytically tractable. While our methods are presented in a very general manner, in Section~\ref{sec:analysis} we examine in detail two multistate dynamical processes specified by different rate matrix functions $\mathbf{F}_{\mathbf{m}}$. Our analysis illustrates how our approach can give a  deep understanding of  multistate dynamical processes on a network and an appreciation for the various levels of accuracy that are required for different dynamical systems. 

The paper is laid out as follows. In Section~\ref{sec:multistate_dynamics}, we review multistate dynamical process models from the literature. The general approximation frameworks are developed in Section~\ref{sec:approximation_frameworks}, while two specific multistate dynamical processes are examined in Section~\ref{sec:analysis}. We conclude in Section~\ref{sec:conclusions}.

\section{Multistate models of dynamical processes}
\label{sec:multistate_dynamics}

Multistate dynamical processes are prominent in the field of epidemiology, where compartmental models are widely used to model the progression of diseases through a population. The simplest epidemiological compartmental models are the susceptible-infected (SI), susceptible-infected-susceptible (SIS) and susceptible-infected-recovered (SIR) models~\cite{kermack1927contribution}, but these models are known to  have limited utility in modelling real epidemics~\cite{pastor2015epidemic}. Instead, more complex multistate compartmental models are used to model and forecast real epidemics. For example the SEIR model and its extended variants, which include an ``exposed'' compartment where individuals have contracted the disease but are not yet infectious, are commonly used to predict the evolution of real epidemics such as  Ebola~\cite{althaus2014estimating,gomes2014assessing,legrand2007understanding,lekone2006statistical} and SARS~\cite{ng2003double,small2005small}.  The SIRS model, where recovered nodes can again become susceptible to infection after an appropriate immunity period, is used as  a model for influenza that can capture the seasonal oscillations of the disease~\cite{dushoff2004dynamical, hooten2010assessing}. In fact, multitudes of other states such as  \emph{vaccinated}~\cite{shulgin1998pulse}, \emph{maternally immune}~\cite{hethcote2000mathematics}, and  \emph{quarantined}~\cite{hethcote2002effects} can be included depending on the problem being modelled; comprehensive overviews of such multi-compartmental models can be found in~\cite{anderson1992infectious,hethcote2000mathematics}. Note that, in general, the analysis of such models assumes mean-field or homogeneous-mixing assumptions, where every node can make contact with every other node~\cite{althaus2014estimating,anderson1992infectious,gomes2014assessing,hethcote2000mathematics,kermack1927contribution,lekone2006statistical,shulgin1998pulse}; however, these assumptions are very often too simple for real populations, which have highly complex network structures.  

It is important to understand the dynamics of multistate epidemic processes---and their interplay with the network on which they spread---in order to gain insights for the prediction and control of real diseases. One particularly important feature of the dynamics is the so-called \emph{epidemic threshold}, a critical point in the dynamics which separates the equilibrium condition of a disease-free network (where no node is infected) from the sustained endemic state\footnote{Strictly speaking, this definition of the epidemic threshold only applies to the idealised case of networks with an infinite number of nodes. In finite networks, the disease will always die out eventually; in this case, the epidemic threshold has been defined~\cite{draief2006thresholds} such that below the threshold the disease dies out exponentially fast, while above the threshold it dies out logarithmically slowly.}. Recently, certain authors have devised generalized descriptions of multistate epidemic processes and analyzed their behaviour and epidemic thresholds on complex networks; such works include that of Lin \emph{et al.}~\cite{lin2014modelling}, Guo \emph{et al.}~\cite{guo2012global}, and the S$^*$I$^2$V$^*$ model of Prakash \emph{et al.}~\cite{prakash2012threshold}. Also noteworthy is the work of Masuda and Konno~\cite{masuda2006multi} who, despite not presenting a generalized model, examine the epidemic thresholds of a wide variety of multistate epidemic processes using degree-based mean-field approaches~\cite{pastor2015epidemic}. Note at this point that different theoretical approaches can lead to differing predictions of the epidemic threshold, and thus extensive numerical simulation can be essential to gauge the accuracy or limitations of such theories~\cite{mata2015multiple,fennell2016limitations}. 

Recently, much focus has shifted to the area of interacting diseases, where multiple diseases co-exist in a population and where each disease can affect the progression of the other diseases~\cite{chen2013outbreaks,hebert2015complex,sanz2014dynamics}. This work has been motivated by dependancies between real diseases, such as the increased rate of tuberculosis progression in individuals who are infected with HIV~\cite{corbett2003growing}. Interacting disease dynamics are naturally expressed in a multistate setting, where the state of each node encodes whether it is susceptible, infected etc.~for each of the co-existing diseases.  This construction also applies to general multiply-interacting dynamical processes such as, for example, the interplay between the spread of a disease and the spread of awareness or information about the disease~\cite{granell2013dynamical}.  

Because epidemiological models describe the diffusion of contagions through a network, they have also been widely employed to model diffusions in the social and life sciences~\cite{castellano2009statistical}. The Bass model for the diffusion of innovations is a well known variant of the SI model that includes external influences, and multistate extensions such as the four-state model of Mellor \emph{et al.}~\cite{mellor2015influence} have been proposed to address limitations that can arise because of the simplicity of the original Bass model. Similarly, multistate contagion models have been used to realistically model a variety of social phenomena, such as Xiong \emph{et al.'s} four-state model of information spread~\cite{xiong2012information} and the four-state fanaticism model of Castillo \emph{et al.}~\cite{castillo2003models,stauffer2007can}. The general multi-stage model of Krapivsky and Redner~\cite{krapivsky2011reinforcement} models the spread of innovations and fads, where the multiple stages represent the social reinforcement required before individuals eventually adopt a behaviour, while de Arruda \emph{et al.}~\cite{de2016unifying} developed a unifying approach for rumour and disease spreading. 

Epidemiological models are not sufficient however to capture all kinds of human interaction behaviour, and several other classes of models exist. One of the most widely studied models of social dynamics in the statistical physics literature is the voter model~\cite{castellano2009statistical,sood2005voter}, which models the adoption of opinions by members of a population through the mechanism of imitation. In its original binary-state form, the voter model eventually leads to consensus within a finite population, where every individual eventually adopts the same opinion~\cite{liggett1985particle}. However, real social dynamics often display richer phenomena, such as social segregation~\cite{schelling1971dynamic}, and so multistate variations of the voter model have been introduced that can account for such complex behaviours. The Leftists, Centrists and Rightists in the model of Vazquez \emph{et al.}~\cite{vazquez2003constrained,vazquez2004ultimate} produce spatially heterogeneous equilibrium states where clusters of Leftists co-exist alongside clusters of Rightists. Other multistate voter models produce sustained meta-stable states of spatial heterogeneity before relaxing to consensus: examples of such models include the four-state confident voting model of Volovik and Redner~\cite{volovik2012dynamics}, the noise reduced voter model of~\cite{dall2007effective} and the three-state $AB$ model of Castell{\'o} \emph{et al.}~\cite{castello2009consensus}. Similarly, in some multistate voter models consensus is never reached, such as the model of Volovik~\emph{et al.}~\cite{volovik2009dynamics}, which includes noise terms that allow for both the suppression or the equalization of opinions. Another multistate model that is well-studied in the statistical physics literature is the Potts model, a multistate extension of the Ising model which, although originally formulated as a model of ferromagnets, has been used to model  cellular dynamics and opinion dynamics~\cite{castellano2009statistical}.

Finally, we discuss threshold models, models of diffusion that fundamentally differ from the voter and epidemiological-type models~\cite{watts2002simple}. These models describe ``complex contagions'', where numerous exposures to an innovation are required before an agent adopts~\cite{centola2007complex}. Specifically, individual nodes in a threshold model will only adopt a behaviour if the fraction of their neighbours who have previously adopted is above a certain threshold. Traditional binary-state threshold models have been extensively studied, and recently multistate threshold models been proposed. Melnik \emph{et al.}~\cite{melnik2013} introduced a ``progressive'' three-state model, where agents have two thresholds, while Kuhlman and Mortviet~\cite{kuhlman2015limit} extended this formalism, allowing an arbitrary number of states and transitions from each state to every other state. A similar non-linear model is the multistate majority rule model~\cite{chen2005consensus}, where every node has a multitude of opinions and nodes change their opinion depending on the majority opinion of groups to which they belong.
 
The models discussed above, although stemming from a diverse range of disciplines with different motivating questions, can all be represented as members of a common mathematical class. In each model, nodes are in one of a discrete number $n$ of states, which we can generically denote as $\{0,1,\dots,n-1\}$ (Fig. 1). Nodes can dynamically change from their current state to another state, and do so at a stochastic rate $F_{\mathbf{m}}(i \rightarrow j)$ that depends on both their own state and on the states of their neighbours, encoded in the vector $\mathbf{m} = \{m_0,m_1,\dots,m_{n-1}\}$ where $m_l$ is the number of neighbours of a node in state $l$. We note the important fact that the rate functions that we study here are independent of time $t$; the dynamical processes that we study here are therefore memoryless or Markovian.

\section{Approximation Frameworks}
\label{sec:approximation_frameworks}
In this section we develop approximation frameworks for the analysis of multistate stochastic dynamical processes, by generalizing the methods for binary-state dynamics of Refs.~\cite{gleeson2011high,gleeson2013}. 
We consider  undirected, unweighted networks with a given degree distribution $p_k$, where $p_k$ is the probability that a randomly chosen node in the network has degree $k$.  We assume that the number $N$ of nodes in the network is very large (taking the $N\to \infty$ limit) and that the networks are  maximally random subject to the constraint of the degree distribution, e.g.,  networks drawn from the configuration-model ensemble \cite{newman2010networks}. Such networks are useful in studying how network connectivity (i.e., the distribution of degrees) affects dynamical processes taking place on  networks; however, they do not possess degree-degree correlations,  transitivity, or other structural properties that are characteristic of real-world networks \cite{newman2010networks}.

Each node in the network can be in one of $n$ states
and nodes
 change state from their current state $i$ to another state $j$ at a rate $F_{\bf{m}}(i \rightarrow j)$ (which can be zero in the case that a transition $i\rightarrow j$ is not possible).
The rate functions $F_{\bf{m}}(i \rightarrow j)$, for $0\leq i,j \leq n-1$ (or, equivalently, the rate matrix function $\mathbf{F}_{\mathbf{m}}$), fully define the continuous-time Markovian dynamical process. In Section~\ref{chpt6:sec2:MSframeworks:AME} we present the approximate master equations (AME), followed by the pair approximation (PA) framework in Section~\ref{chpt6:sec2:MSframeworks:PA} and finally the mean-field (MF) framework in Section~\ref{chpt6:sec2:MSframeworks:MF}. The reason for this order is that the PA equations can be deduced from the AME equations by making appropriate simplifying assumptions, and similarly the MF equations can be deduced from the PA equations under further assumptions.

\subsection{Approximate Master Equation}
\label{chpt6:sec2:MSframeworks:AME}

The approximate master equation (AME) is a theoretical framework for studying  dynamical processes that has been shown to reproduce a  range of binary-state dynamics to a high level of accuracy~\cite{fennell2014analytical,gleeson2011high,gleeson2013,marceau2010adaptive}. We define as $x_{k,\bf{m}}^i(t)$ the variables of the multistate AME, where $x_{k,\bf{m}}^i(t)$ is the expected fraction of $k$-degree nodes in the network that are in state $i$ and have $\bf{m}$ neighbours in each of the various states at time $t$. These variables are defined for all states $0\leq i\leq n-1$, for all possible degree classes $k_{\text{min}}\leq k \leq k_{\text{max}}$ and --- for each degree class $k$ --- all values of $\bf{m}$ such that $\sum_{l=0}^{n-1}m_l = k$. From these variables, one can obtain various macroscopic quantities related to the evolution of the dynamics on the network, such as the expected fraction $\rho^i(t)$ of the population in  state $i$ at time $t$:
\begin{equation}
    \rho^i(t) = \Big\langle\sum_{|\mathbf{m}|=k}x_{k,\bf{m}}^i(t)\Big\rangle_k.
\end{equation}
Here $\sum_{|\mathbf{m}|=k}$ is the sum over all values of $\bf{m}$ such that $\sum_{l=0}^{n-1}m_l = k$ and $\langle \cdot \rangle_k = \sum_{k=0}^{\infty}p_k\cdot\,\,$  symbolizes averaging over the degree distribution $p_k$.

\begin{figure}
    \centering
    \includegraphics[width=0.9\textwidth]{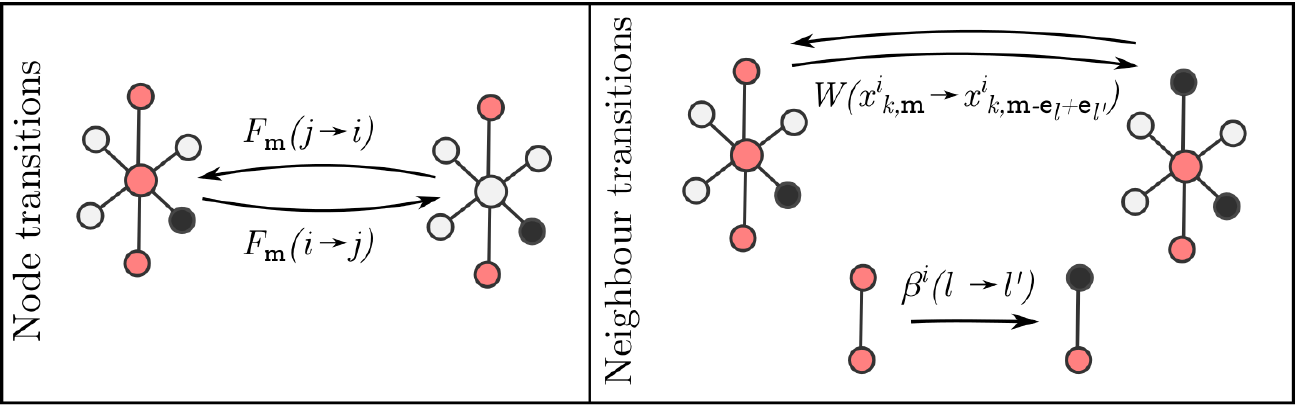}
    \caption{Transitions between classes in the AME formalism. Nodes change from one class to another either because they change state (left) or one of their neighbours changes state (right). The node transitions are fully specified by the rate functions $F_{\mathbf{m}}(i \rightarrow j)$. The neighbour transition rates $W(x_{k,\bf{m}}^i \rightarrow x_{k,\bf{m}-\bf{e_l}+\bf{e_{l'}}}^i)$ are approximated by link transition rates $\beta^i(l\rightarrow l')$ as given by Eqs.~\eqref{eq:transition_approx} and \eqref{eq:def:beta_multistate}.}
    \label{fig:transitions}
\end{figure}

To derive the evolution equations that describe how the variables $x_{k,\bf{m}}^i(t)$ change over time, we consider all ways in which nodes can leave and enter the $x_{k,\bf{m}}^i$ class. These changes are illustrated in the schematic of Fig.~\ref{fig:transitions}. On one hand, nodes in the $x_{k,\bf{m}}^i$ class will \emph{leave} that class because either the node itself changes state (a ``node transition'') or because one of its neighbours changes state, and thus the vector $\mathbf{m}$ changes (a ``neighbour transition''). In an (infinitesimally small) time interval $\Delta t$, a $x_{k,\bf{m}}^i$ node will change  from state $i$ to state $j$ with probability  $F_{\bf{m}}(i \rightarrow j)\Delta t$, and so the expected fraction of  nodes in the $x_{k,\bf{m}}^i$ class that change to state $j$ is $x_{k,\bf{m}}^i(t)F_{\bf{m}}(i \rightarrow j)\Delta t$. Accounting for all states $j\neq i$, the expected fraction of nodes that leave the $x_{k,\bf{m}}^i$ class because they themselves change state from state $i$ is
\begin{equation}
    \sum_{j\neq i}x_{k,\bf{m}}^i(t)F_{\bf{m}}(i \rightarrow j)\Delta t .
    \label{eq:def:total_node_trans_rate_out}
\end{equation}
In addition, nodes can leave the $x_{k,\bf{m}}^i$ class because their neighbours change state. The probability that a neighbour of such a node that is currently in state $l$  changes to state $l'$ in the infinitesimal time interval $\Delta t$ is given by
\begin{equation}
    W(x_{k,\bf{m}}^i \rightarrow x_{k,\bf{m}-\bf{e_l}+\bf{e_{l'}}}^i)\Delta t
    \label{eq:def:neighbour_transition_rate}
\end{equation}
where $W(x_{k,\bf{m}}^i \rightarrow x_{k,\bf{m}-\bf{e_l}+\bf{e_{l'}}}^i)$ is a neighbour transition rate and $\bf{e_{l}}$ (resp. $\bf{e_{l'}}$) is the standard unit basis vector which is zero everywhere except at position $l$ (resp. $l'$) (and so $\mathbf{m}-\mathbf{e_l}+\mathbf{e_{l'}} = \{\dots,m_l-1,\dots, m_{l'}+1, \dots\}$). It is in these neighbour transition rates that we make 
the AME approximation in order to create a closed system of equations. This approximation is the assumption that the transition rate of a given neighbour of a node is independent of the states of all other neighbours of the node. Thus, the rate at which a neighbour of a state $i$ node will change from state $l$ to state $l'$ is assumed to be equal to  the rate at which a link of type $(i)$---$(l)$ changes to one of type $(i)$---$(l')$, averaged over the whole network (see Fig.~\ref{fig:transitions}). We denote this link transition rate as $\beta^{i}(l\rightarrow l')$, and thus we have that the neighbour transition rate of Eq.~\eqref{eq:def:neighbour_transition_rate} is approximated as 
\begin{equation}
    W(x_{k,\bf{m}}^i \rightarrow x_{k,\bf{m}-\bf{e_l}+\bf{e_{l'}}}^i) = m_l\beta^{i}(l\rightarrow l').
    \label{eq:transition_approx}
\end{equation}
The link transition rate $\beta^{i}(l\rightarrow l')$ is formulated by calculating a global average over the whole network, as follows. The expected number of links of type $(i)$---$(l)$ in the network at time $t$ is $N\langle \sum_{|\mathbf{m}|=k}m_ix_{\bf{m}}^l(t)\rangle_k$, and the expected number of these $(i)$---$(l)$ links that change to $(i)$---$(l')$ in the infinitesimal time interval $\Delta t$ is $N\langle \sum_{|\mathbf{m}|=k}m_ix_{\bf{m}}^l(t)F_{\bf{m}}(l\rightarrow l')\Delta t\rangle_k$. Then the probability $\beta^{i}(l\rightarrow l')\Delta t$ that a link of type $(i)$---$(l)$ changes to a link of type $(i)$---$(l')$ is the ratio of these two quantities and so $\beta^{i}(l\rightarrow l')$ is given by
\begin{equation}
    \beta^{i}(l\rightarrow l') = \frac{\langle \sum_{|\mathbf{m}|=k}m_ix_{k,\mathbf{m}}^l(t)F_{\bf{m}}(l\rightarrow l')\rangle_k}{\langle \sum_{|\mathbf{m}|=k}m_ix_{k,\mathbf{m}}^l(t)\rangle_k}.
    \label{eq:def:beta_multistate}
\end{equation}
Accounting for each possible transition $l\rightarrow l'$ and each of the neighbours of a node in the $x_{k,\bf{m}}^i$ class gives the expected fraction of nodes that leave the class during the  $\Delta t$  time interval because their neighbours change state:
\begin{equation}
    \sum_{l=0}^{n-1}\sum_{l'\neq l}x_{k,\bf{m}}^i(t)m_l\beta^{i}(l\rightarrow l')\Delta t.
    \label{eq:def:total_neighb_trans_rate_out}
\end{equation}

On the other hand, nodes in a different class will \emph{enter} the $x_{k,\bf{m}}^i$ class because either their state changes to $i$ or the state of one of their neighbours changes. A node in the $x_{k,\bf{m}}^j$ class will enter the $x_{k,\bf{m}}^i$ class if its state changes from $j$ to $i$; this occurs with probability $F_{\bf{m}}(j \rightarrow i)\Delta t$ and if we consider all classes $j \neq i$, then the expected fraction of nodes that enter the $x_{k,\bf{m}}^i$ class because they change state to $i$ is given by
\begin{equation}
    \sum_{j\neq i}x_{k,\bf{m}}^j(t)F_{\bf{m}}(j \rightarrow i)\Delta t.
    \label{eq:def:total_node_trans_rate_in}
\end{equation}
In terms of neighbour transitions, a node in the $x_{k,\bf{m}-\bf{e_l}+\bf{e_{l'}}}^i$ class can enter the $x_{k,\bf{m}}^i$ class if one of its neighbours in state $l'$ changes to state $l$. This will occur with probability $(m_{l'}+1)\beta^{i}(l'\rightarrow l)\Delta t$---where $\beta^{i}(l'\rightarrow l)$ is  defined as in Eq.~\eqref{eq:def:beta_multistate}---and so the expected fraction of nodes that enter the $x_{k,\bf{m}}^i$ class during a $\Delta t$ time interval because of a neighbour changing state is
\begin{equation}
    \sum_{l=0}^{n-1}\sum_{l'\neq l}x_{k,\bf{m}-\bf{e_l}+\bf{e_{l'}}}^i(t)(m_{l'}+1)\beta^{i}(l'\rightarrow l)\Delta t.
    \label{eq:def:total_neighb_trans_rate_in}
\end{equation}
Thus, we have quantified all ways in which the size of the  $x_{k,\bf{m}}^i(t)$ class can change in an infinitesimally small time interval $\Delta t$. (Note that in continuous-time Markov processes at most one transition can occur in a sufficiently small time interval; multiple transitions occur with probabilities that are of order $\left(\Delta t\right)^2$ as $\Delta t \to 0$, and are therefore negligible when we derive the differential equations below).  

The value $x_{k,\bf{m}}^i(t+\Delta t)$ at the end of the $\Delta t$ time interval is given by
\begin{equation}
    x_{k,\bf{m}}^i(t+\Delta t) = x_{k,\bf{m}}^i(t) - leave + enter,
    \label{eq:change}
\end{equation}
where the $leave$ term of Eq.~\eqref{eq:change} is comprised of the changes given by Eqs.~\eqref{eq:def:total_node_trans_rate_out} and \eqref{eq:def:total_neighb_trans_rate_out} and the $enter$ term is comprised of the changes given by Eqs.~\eqref{eq:def:total_node_trans_rate_in} and \eqref{eq:def:total_neighb_trans_rate_in}. Finally, if we divide Eq.~\eqref{eq:change} by $\Delta t$,  and take the limit $\Delta t \rightarrow 0$, we arrive at the evolution equation for $x_{k,\bf{m}}^i(t)$:
\begin{multline}
    \frac{d}{dt}x_{k,\bf{m}}^i = -\sum_{j\neq i}F_{\bf{m}}(i \rightarrow j)x_{k,\bf{m}}^i + \sum_{j\neq i}F_{\bf{m}}(j \rightarrow i)x_{k,\bf{m}}^j \\
    - \sum_{l=0}^{n-1}\sum_{l'\neq l}m_l\beta^{i}(l\rightarrow l')x_{k,\bf{m}}^i
    +  \sum_{l=0}^{n-1}\sum_{l'\neq l}(m_{l'}+1)\beta^{i}(l'\rightarrow l)x_{k,\bf{m}-\bf{e_l}+\bf{e_{l'}}}^i
    \label{eq:def:master_equation_multistate}
\end{multline}
Equation~\eqref{eq:def:master_equation_multistate}, for $0\leq i\leq n-1$ and for all values of $\bf{m}$ and all degree classes $k_{\text{min}}\leq k \leq k_{\text{max}}$, forms a closed set of equations that describe the evolution of the system from known initial conditions $x_{k,\bf{m}}^i(0)$. The number of equations in this system can be calculated by considering all possible combinations of $\mathbf{m}$ for a particular value of $k$. Such $\mathbf{m}$ variables must satisfy $\sum_{l=0}^{n}m_l=k$. This combinatorial problem can be directly mapped to an urn problem, where there are $n$ different colored balls in an urn and one must draw $k$ balls from the urn \emph{with} replacement~\cite{feller2008introduction}. The number of different possible draws here is $\binom{k+n-1}{k}$, and if we consider all values of $i$ and $k$ then the size of the AME system of equations is
\begin{equation}
    n\sum_{k=k_{\text{min}}}^{k_{\text{max}}}\binom{n+k-1}{k}.
    \label{eq:AME_size}
\end{equation}
Equation~\eqref{eq:AME_size} scales super-linearly with both $m$ and the maximum degree $k_{\text{max}}$ of the degree distribution. In many cases, this system size can prove too large to allow for any meaningful analysis, and so in subsequent sections we define simpler systems, at the cost of further approximation.

\subsection{Pair Approximation Framework}
\label{chpt6:sec2:MSframeworks:PA}

The AME makes the assumption that the transition rate of a neighbour of a given node is independent of the states of all other neighbours of the node. The pair approximation (PA) goes a step further by assuming that the \emph{state} of a neighbour of a given node is independent of the states of all other neighbours of the node. Dynamical correlations between a node and each individual nearest neighbour are considered but the states of the neighbours of the node are assumed to be independent. We define as $x_k^i(t)$ and $q_k^{i\rightarrow j}(t)$ the variables of the multi-state PA, where $x_k^i(t)$ is the fraction of $k$-degree nodes in state $i$ at time $t$ and $q_k^{i\rightarrow j}(t)$ is the probability at time $t$ that a randomly chosen link emanating from a $k$-degree node in state $i$ leads to a node in state $j$. Under the PA assumption that the states of the neighbours of a node are independent, the probability of having $\mathbf{m}$ neighbours in each of the various states is multinomially distributed. If we denote by $\text{Mult}_{k,i}(\mathbf{m})$ the probability that a $k$-degree node in state $i$ has $\bf{m}$ neighbours in the various different states at time $t$, then $\text{Mult}_{k, i}(\mathbf{m})$ is given by
\begin{equation}
    \text{Mult}_{k, i}(\mathbf{m}) = \frac{k!}{m_0!\dots m_{n-1}!}(q_k^{i\rightarrow 0}(t))^{m_0}\dots(q_k^{i\rightarrow m-1}(t))^{m_{n-1}}.
    \label{eq:def:Mult}
\end{equation}
By making the PA independence assumption, the AME variables $x_{k,\bf{m}}^i(t)$ reduce to
\begin{equation}
    x_{k,\bf{m}}^i(t) = x_{k}^i(t)\text{Mult}_{k, i}(\mathbf{m}).
    \label{eq:def:PA_ansatz}
\end{equation}
Note that the sum of the multinomial probabilities of Eq.~\eqref{eq:def:Mult} over all values of $\mathbf{m}$ with $|\mathbf{m}|=k$ is 1, and performing this sum on both sides of Eq.~\eqref{eq:def:PA_ansatz} gives a direct relationship between the PA variable $x_{k}^i(t)$ and the AME variables $x_{k,\bf{m}}^i(t)$:
\begin{equation}
    x_{k}^i(t) = \sum_{|\mathbf{m}|=k}x_{k,\bf{m}}^i(t).
    \label{eq:PA_to_AME_x}
\end{equation}
By differentiating both sides of Eq.~\eqref{eq:PA_to_AME_x} with respect to time, and inserting the PA ansatz of Eq.~\eqref{eq:def:PA_ansatz} into the right hand side of the evolution Eq.~\eqref{eq:def:master_equation_multistate} for $x_{k,\bf{m}}^i$, we arrive at the PA evolution equation for $x_{k}^i(t)$:
\begin{multline}
    \frac{d}{dt}x_{k}^i = -\sum_{j\neq i}x_{k}^i\sum_{|\mathbf{m}|=k}F_{\textbf{m}}(i \rightarrow j) \text{Mult}_{k, i}(\mathbf{m}) + \sum_{j\neq i}x_{k}^j\sum_{|\mathbf{m}|=k}F_{\textbf{m}}(j \rightarrow i)\text{Mult}_{k, j}(\mathbf{m}).
    \label{eq:def:PA_evolution_node}
\end{multline}

The evolution equations for the other variables of the PA system, $q_k^{i\rightarrow j}(t)$, are constructed from the AME equations in the following manner. In the AME, the expected number of links of type $(i)_k$---$(j)$ at time $t$ (i.e., links emanating from $k$-degree nodes in state $i$ that lead to  a node in state $j$) is $N\sum_{|\mathbf{m}|=k}m_jx_{k,\mathbf{m}}^i(t)$. By multiplying the PA ansatz of Eq.~\eqref{eq:def:PA_ansatz} by $Nm_j$, summing over all values with  $|\mathbf{m}|=k$ and using the fact that $\sum_{|\mathbf{m}|=k}m_j\text{Mult}_{k,i}(\mathbf{m})=kq_k^{i\rightarrow j}(t)$, we obtain
\begin{equation}
     N\sum_{|\mathbf{m}|=k}m_jx_{k,\mathbf{m}}^i(t) = Nx_k^i(t)kq_k^{i\rightarrow j}(t)
     \label{eq:PA_q}
\end{equation}
Differentiating both sides of Eq.~\eqref{eq:PA_q} with respect to time, cancelling the common factor $N$ and rearranging gives
\begin{equation}
     \frac{d}{dt}q_k^{i\rightarrow j} = - \frac{1}{x_k^i}\left(\frac{dx_k^i}{dt}q_k^{i\rightarrow j} - \sum_{|\mathbf{m}|=k}\frac{m_j}{k}\frac{dx_{k,\mathbf{m}}^i}{dt}\right).
     \label{eq:PA_link_unspecified}
\end{equation}
Inserting into Eq.~\eqref{eq:PA_link_unspecified} the expressions for $dx_k^i/dt$ and $dx_{k,\mathbf{m}}^i/dt$ from Eqs.~\eqref{eq:def:PA_evolution_node} and \eqref{eq:def:master_equation_multistate}, respectively, and using the PA ansatz of Eq.~\eqref{eq:def:PA_ansatz} gives the evolution equation for $q_k^{i\rightarrow j}(t)$:
\begin{multline}
     \frac{d}{dt}q_k^{i\rightarrow j} = \\ \sum_{|\mathbf{m}|=k}\left(q_k^{i\rightarrow j} - \frac{m_j}{k}\right)\left(\sum_{l\neq i}F_{\mathbf{m}}(i \rightarrow l)\text{Mult}_{k,i}(\mathbf{m}) - \frac{x_k^l}{x_k^i}F_{\mathbf{m}}(l \rightarrow i)\text{Mult}_{k,l}(\mathbf{m})\right) \\ + \sum_{l=0}^{n-1}\Big(q_k^{i\rightarrow l}\beta^i(l\rightarrow j)-q_k^{i\rightarrow j}\beta^i(j\rightarrow l)\Big).
     \label{eq:def:PA_evolution_link}
\end{multline}
The link transition rates $\beta^i(l\rightarrow j)$ for the PA equations~\eqref{eq:def:PA_evolution_link} are obtained by inserting the PA ansatz into the AME link transition rates of Eq.~\eqref{eq:def:beta_multistate}, yielding
\begin{equation}
    \beta^{i}(l\rightarrow j) = \frac{\langle x_{k}^l(t) \sum_{|\mathbf{m}|=k}m_i\text{Mult}_{k,l}(\mathbf{m})F_{\bf{m}}(l\rightarrow j)\rangle_k}{\langle x_k^l(t)kq_k^{l \rightarrow i}(t) \rangle_k}.
    \label{eq:def:beta_PA}
\end{equation}
Thus we have completely described the PA framework, which is comprised of Eqs.~\eqref{eq:def:PA_evolution_node} and \eqref{eq:def:PA_evolution_link} for $0\leq i,j\leq n-1$ and for $k_{\text{min}}\leq k \leq k_{max}$. This is a closed set of $(n^2+n)(k_{\text{max}} - k_{\text{min}} +1)$ equations that describes the evolution of the system from initial conditions $x_k^i(0)$ and $q_k^{i\rightarrow j}(0)$. Importantly, the size of the system scales linearly with the range $k_{\text{max}} - k_{\text{min}}$ of possible degrees, which is relatively modest compared to the size of the AME system of Eq.~\eqref{eq:def:master_equation_multistate}. This is especially relevant when studying networks with heterogeneous degree distributions, i.e., with large values of $k_{\text{max}} - k_{\text{min}}$.

\subsection{Mean-Field Framework}
\label{chpt6:sec2:MSframeworks:MF}

In the mean-field (MF) approximation scheme it is assumed that the states of each node in the network are independent. Dynamical correlations that may exist between a node and its nearest neighbours are, as a result, neglected. To arrive at the MF equations from the PA framework, we note that the link probabilities $q_k^{i\rightarrow j}(t)$ are independent of $i$ under the MF assumption. Thus for all values of $i$, $q_k^{i\rightarrow j}(t)$ is replaced by $\omega^j(t)$, the mean-field probability that the neighbour of a node is in state $j$ at time $t$. The value of  $\omega_j(t)$ is approximated directly from the $x_{k}^i(t)$ terms by a global average of node states over the entire network, given by
\begin{equation}
    \omega^j(t) = \sum_{k = 0}^{\infty}\frac{k p_{k}}{z}x_{k}^j(t),
    \label{eq:def:MF_ansatz}
\end{equation}
where $z = \langle k \rangle$  is the average degree of the network. In Eq.~\eqref{eq:def:MF_ansatz}, $kp_{k}/z$ is the probability that a neighbour of a node has degree $k$ in the configuration network model, while $x^j_k(t)$ is the probability that such a neighbour is in state $i$. Summing over all possible values of neighbour degrees $k$ gives the desired probability $\omega^j(t)$. The mean-field ansatz of Eq.~\eqref{eq:def:MF_ansatz} can be inserted into the PA evolution equation for $x_k^i$ to give the MF evolution equation for $x_{k}^i(t)$:
\begin{multline}
    \frac{d}{dt}x_{k}^i = -\sum_{j\neq i}x_{k}^i \sum_{|\mathbf{m}|=k} \text{Mult}_{k}(\mathbf{m}) F_{\textbf{m}}(i \rightarrow j) +\sum_{j\neq i}x_{k}^j \sum_{|\mathbf{m}|=k} \text{Mult}_{k}(\mathbf{m})F_{\textbf{m}}(j \rightarrow i)
    \label{eq:def:MF_evolution_node}
\end{multline}
where, similar to Eq.~(\ref{eq:def:Mult}), $\text{Mult}_{k}(\mathbf{m})$ is defined as
\begin{equation}
    \text{Mult}_{k}(\mathbf{m}) = \frac{k!}{m_0!\dots m_{n-1}!}(\omega^0)^{m_0}\dots(\omega^{n-1})^{m_{n-1}}.
    \label{eq:def:Mult_2}
\end{equation}
Equation~\eqref{eq:def:MF_evolution_node}, for $0 \leq i \leq n-1$ and $k_{\text{min}}\leq k \leq k_{\text{max}}$, is a closed system of $n(k_{\text{max}} - k_{\text{min}} +1)$ equations that describe the evolution of the MF dynamics from initial conditions $x_{k}^i(0)$. Note that the mean-field equations can be represented eloquently in vector form; if we denote by $\mathbf{x}_{k}(t)$ the $n\times 1$ vector whose $i$'th entry is $x^i_{k}(t)$, then the evolution equation for $\mathbf{x}_{k}(t)$ is given by
\begin{equation}
    \frac{d}{dt}\mathbf{x}_k = -\sum_{|\mathbf{m}|=k}\left(\mathbf{R}_{\mathbf{m}} - \mathbf{F}_{\mathbf{m}}^{\text{T}}\right)\text{Mult}_{k}(\mathbf{m})\mathbf{x}_k,
\end{equation}
where $\mathbf{F}_{\mathbf{m}}^T$ is the transpose of the transition rate matrix and $\mathbf{R}_{\mathbf{m}}$ is the diagonal matrix with elements $(\mathbf{R}_{\mathbf{m}})_{ii}=\sum_{j=1}^nF_{\mathbf{m}}(i\rightarrow j)$.

To recap, we have derived the AME, PA and MF approximation frameworks by making a series of progressively stronger assumptions. The AME assumes that the transition rate of the neighbour of a given node is independent of the states of all other neighbours of the node. Note also that by closing the AME system of equations we explicitly assume independence between a node and all other nodes in the network beyond its nearest neighbours (i.e., an absence of long range correlations). The PA brings the level of approximation a step further by assuming that the \emph{state} of the neighbour of a node is independent of the states of all other neighbours of the node, while the MF makes the strongest possible assumption with the ansatz that the state of every node in the network is independent. 

\subsection{Numerical solvers}

In Sections \ref{chpt6:sec2:MSframeworks:AME} through \ref{chpt6:sec2:MSframeworks:MF} we have presented the AME, PA and MF theoretical frameworks and have seen that the number of equations in the approximation frameworks markedly decreases as simplifying assumptions are made. While the AME scales super-linearly in both the number of states $n$ and the maximum degree $k_{\text{max}}$, the PA scales linearly in the range $k_{\text{max}} - k_{\text{min}}$ of the degree distribution while the MF scales linearly in both $n$ and $k_{\text{max}} - k_{\text{min}}$. The latter frameworks can be attractive for exploring dynamical processes, as their relatively small system size lends itself to analytical study. However, the simplifying assumptions employed to arrive at such frameworks may result in a significant loss of accuracy or even a complete failure to capture the dynamics, and so there are cases when the higher accuracy approximation schemes, although possibly not solvable analytically, can give valuable qualitative and quantitative (through numerical solution) insights into the dynamics. For the cases where analytical solutions are not attainable, we make freely available a MATLAB/Octave package to numerically solve the systems of equations~\cite{website:MultiAMEcode}, which has been  optimized to deal with the large number of equations that can occur particularly in the case of the AME.

\section{Analysis}
\label{sec:analysis}
In this section, we illustrate the power of the approximation frameworks introduced in Section~\ref{sec:approximation_frameworks} in understanding multistate dynamical processes. We first illustrate our approach with the study of co-evolving epidemiological dynamics,  where two epidemic processes interact with each other on a network. Our equations reveal the rich equilibrium behaviour of these dynamics, and the interplay between the connectivity of the network and the infectivity parameters of the diseases. Following this we present a dynamical process from the physical sciences which displays jamming or segregation at equilibrium, and we demonstrate the necessity of high accuracy approximations for capturing such complex behaviour.

\subsection{Cooperative disease dynamics}
\label{sec:CoSIS}

\begin{figure}
    \centering
    \includegraphics[width=0.9\columnwidth]{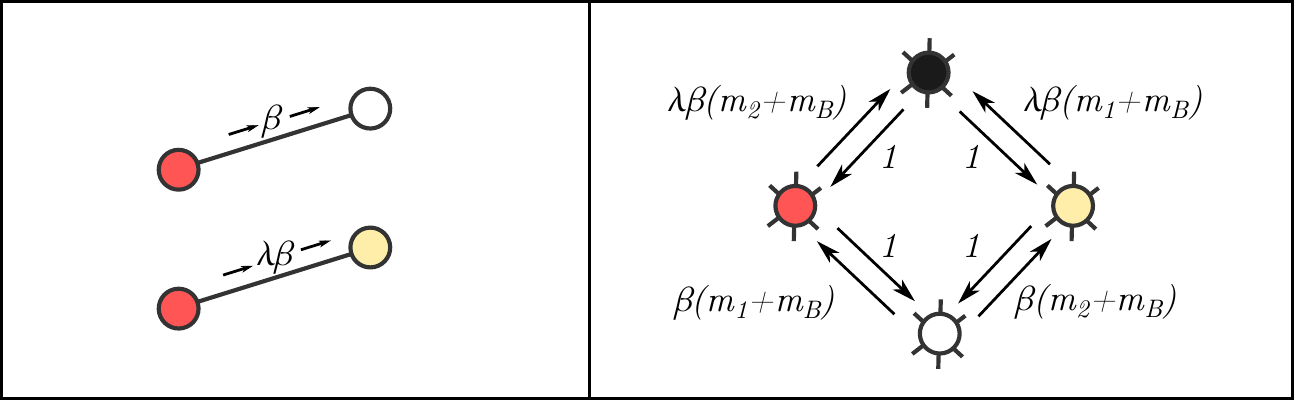}
    \caption{Schematic of the transition rates in the cooperative SIS model. In this model, individuals can be in one of four states -- susceptible (white), infected with disease 1 (red), infected with disease 2 (yellow), and infected with both diseases (black). Left: A disease will be transmitted from a node with one disease to a susceptible node at a rate $\beta$, while it will be transmitted to a node that is infected with the other disease at an accentuated rate $\lambda\beta$. Right: The transition rates between the various states. Here, $m_1, m_2$ and $m_B$ are the number of neighbours of the nodes with, respectively, disease 1 only, disease 2 only, and with both diseases.}
    \label{fig:Co_SIS_schematic}
\end{figure}

The study of cooperative diseases, where one disease can facilitate the spread of another disease, is an area that is gaining much scientific attention~\cite{chen2013outbreaks,janssen2016first,hebert2015complex,sanz2014dynamics}, motivated by the dependencies between real diseases such as HIV and Tuberculosis~\cite{corbett2003growing}. The model we study here is a relatively simple model of two cooperative SIS processes which we label disease 1 and disease 2. 
In their individual behaviours (i.e., in the absence of the other disease), the diseases are identical: Individuals infected with a single disease will transmit that disease to each of their neighbours at a rate $\beta$, while infected individuals recover at rate 1. Interaction between the two diseases occurs via the mechanism whereby individuals already infected with one disease are more likely to become infected with the other disease.
A node infected with one disease will contract the other disease at a rate $\lambda \beta$ from each of its neighbours that is infected with the other disease, where $\lambda \geq 1$  is an accentuation parameter  quantifying the increased vulnerability of infected nodes relative to healthy nodes. The transitions in the model are illustrated in the schematic of Fig.~\ref{fig:Co_SIS_schematic}. This specific model has previously been studied analytically in the case of fully mixed populations~\cite{chen2016phase,chen2013outbreaks,janssen2016first}, while also receiving limited analytical treatment in the case of clustered networks and their tree-like equivalents~\cite{hebert2015complex}. We also note that it is a special case of the very general ten-parameter model of co-evolving SIS processes introduced by Sanz \emph{et al.}~\cite{sanz2014dynamics}. The aim of our work here is to gain a full understanding of the dynamics of the model and the interplay between the disease parameters $\beta$ and $\lambda$ and the network connectivity characteristics i.e., the degree distribution $p_k$.

In the absence of interaction (i.e., if $\lambda = 1$), each disease independently behaves as a traditional binary-state SIS process. For such processes, the critical value of $\beta$ at which an endemic disease state persists in the population is given by degree-based mean-field theory~\cite{pastor2001epidemic} as
\begin{equation}
    \beta_c = \frac{z}{\langle k^2\rangle},
    \label{eq:ET1disease}
\end{equation}
where $\langle k^2\rangle$ is the  second moment of the degree distribution. The interesting question, then, is the effect that interaction between the two diseases will have on the epidemic threshold $\beta_c$ of each of the diseases. We examine this in detail through our degree-based frameworks with the aim of qualitatively exploring the state space $(\lambda, \beta)$ in terms of the equilibrium behaviour of the interacting diseases.

The co-evolving disease model we have described above can be represented as a four-state model where each node is either susceptible ($S$), infected only with disease 1 ($I_1$), infected only with disease 2 ($I_2$) or infected with both diseases ($B$). The transitions of a node between states are encoded in the rate functions $F_{\mathbf{m}}(i \rightarrow j)$, where $\mathbf{m} = \{m_{S}, m_{1},m_2,m_B\}$ is  the number of neighbours a node has in the various  states. These transitions are illustrated in Fig.~\ref{fig:Co_SIS_schematic}. Nodes infected with a disease recover from that disease at a rate 1, so $F_{\mathbf{m}}(I_1 \rightarrow S) =     F_{\mathbf{m}}(I_2 \rightarrow S) =    F_{\mathbf{m}}(B \rightarrow I_1) =    F_{\mathbf{m}}(B \rightarrow I_2) = 1$.
On the other hand, susceptible nodes will contract disease 1 from each of their $m_1+m_B$ neighbours that are infected with disease 1 at a rate $\beta$, and disease 2 from each of their $m_2+m_B$ neighbours that are infected with disease 2 at the same rate $\beta$; this gives $    F_{\mathbf{m}}(S \rightarrow I_1) = (m_1+m_B)\beta$ and $F_{\mathbf{m}}(S \rightarrow I_2) = (m_2+m_B)\beta$. Nodes that are already infected with one disease contract the other disease at an accentuated rate $\lambda\beta$, thus $F_{\mathbf{m}}(I_1 \rightarrow B) = (m_1+m_B)\lambda\beta$ and $F_{\mathbf{m}}(I_2 \rightarrow B) = (m_2+m_B)\lambda\beta$. No further transitions are possible and so the transition rate matrix $\mathbf{F}_{\mathbf{m}}$ for this cooperative SIS model is given by
\begin{equation}
  \mathbf{F}_{\mathbf{m}}=
  \bordermatrix{
    \hspace{4pt} & S & I_1 & I_2 & B  \vspace{4pt} \cr
    \vspace{4pt}
    S \hspace{4pt} & 0 & (m_1+m_B)\beta & (m_2+m_B)\beta & 0 & \cr
    \vspace{4pt}
    I_1 \hspace{4pt} & 1 & 0 & 0 & (m_2+m_B)\lambda\beta  \cr
    \vspace{4pt}
    I_2 \hspace{4pt} & 1 & 0 & 0 & (m_1+m_B)\lambda\beta  \cr
    \vspace{4pt}
    B \hspace{4pt} & 0 & 1 & 1 & 0 \cr
  }
  \label{eq:CoSISratemat}
\end{equation}

\begin{figure}
  \includegraphics[width=0.46\textwidth]{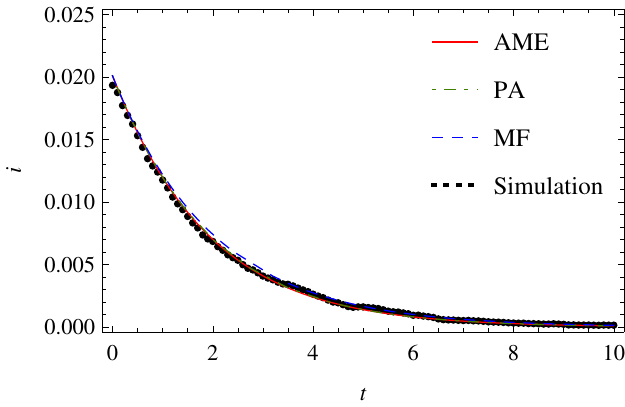} \hspace{10pt}
  \includegraphics[width=0.45\textwidth]{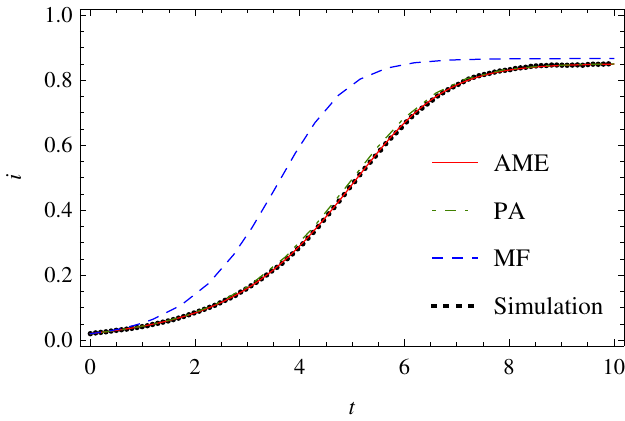}
  \caption{The expected fraction of infected nodes $i(t)$ in a degree-regular network ($z=8$) with $\lambda=2$ and $\beta=0.5\beta_c$ (left) and $\beta=2\beta_c$ (right), where $\beta_c=1/8$ here is the prediction of the critical point for regular SIS dynamics as given by Eq.~\eqref{eq:ET1disease}. Each plot shows $i(t)$ as given by the AME, PA, MF frameworks, as well as $i(t)$ attained through numerical simulations of the dynamics on a network of $10^4$ nodes using a Gillespie algorithm~\cite{gillespie1977}.}
  \label{ref:initcoSIS}
\end{figure}

To begin, we study the evolution of the two diseases on a degree-regular random network, i.e., one in which every node has the same degree $z$, so the degree distribution is  $p_k = \delta_{k,z}$. We analyze the expected fraction of infected individuals $i(t)$ in the population for different values of $\beta$ and $\lambda$, where $i(t)$ includes all nodes with disease 1, disease 2 and both diseases. Figure~\ref{ref:initcoSIS} shows the time evolution of $i(t)$ for two  values of $\beta$, one below and one above the  single-disease epidemic threshold of Eq.~\eqref{eq:ET1disease}, with the accentuation parameter taking the value $\lambda=2$. We plot $i(t)$ from the AME, PA and MF systems of equations as well as $i(t)$ constructed from numerical simulations, which throughout this work will be performed using the Gillespie Algorithm~\cite{gillespie1977,fennell2016limitations}. From Fig.~\ref{ref:initcoSIS}, it is clear that the AME and PA have an excellent level of accuracy, matching the numerical simulations to a high degree of accuracy above and below the critical point. Moreover the MF framework, though deviating from the simulations in the transient regime for $\beta$ above the critical point,
is still quite accurate in the equilibrium regime ($t\to \infty$). Indeed, previous numerical studies have shown that MF frameworks can give highly accurate results for the equilibrium behaviour of SIS dynamics~\cite{mata2015multiple}. We thus employ the MF framework for our analysis because of a) its level of accuracy and b) its relative tractability over the PA and AME frameworks.  

To begin our analysis using the mean-field framework of Eq.~\eqref{eq:def:MF_evolution_node} on degree-regular networks we denote as $x_z^S$, $x_z^1$, $x_z^2$ and $x_z^B$ the fraction of nodes in the network in each of the four states. Note we use the $z$ subscript instead of $k$ as every node in a degree-regular networks has the same degree $k=z$. The evolution equations for $x_z^S$, $x_z^1$, $x_z^2$ and $x_z^B$ are attained by inserting the rate functions $F_{\mathbf{m}}(i \rightarrow j)$ into Eq.~\eqref{eq:def:MF_evolution_node}. In Eq.~\eqref{eq:def:MF_evolution_node}, the rate functions are part of the expression
\begin{equation}
    \sum_{|\mathbf{m}|=z}F_{\mathbf{m}}(i \rightarrow j)\text{Mult}_{z}(\mathbf{m}, \mathbf{\omega}),
    \label{eq:MF_F_Mult_expressions}
\end{equation}
where $\mathbf{\omega}$ in the case of a degree-regular network is simply  $\mathbf{\omega} = \{x_z^S,x_z^1, x_z^2, x_z^B\}$. For the cases when $F_{\mathbf{m}}(i \rightarrow j)=1$, as for the recovery transitions $I_1 \rightarrow S$, $I_2 \rightarrow S$, $B \rightarrow I_1$ and $B \rightarrow I_2$, Eq~\eqref{eq:MF_F_Mult_expressions} becomes
\begin{equation}
    \sum_{|\mathbf{m}|=z}1\times\text{Mult}_{k}(\mathbf{m}, \mathbf{\omega}) = 1,
\end{equation}
because the sum of multinomial probabilities over all possible events is equal to 1. For the infection transitions, we first consider the transition $S \rightarrow I_1$, which occurs at rate $F_{\mathbf{m}}(S \rightarrow I_1)=(m_1+m_B)\beta$. In this case, Eq.~\eqref{eq:MF_F_Mult_expressions} yields
\begin{equation}
    \beta\sum_{|\mathbf{m}|=z}m_1\text{Mult}_{z}(\mathbf{m}, \mathbf{\omega}) + \beta\sum_{|\mathbf{m}|=z}m_B\text{Mult}_{z}(\mathbf{m}, \mathbf{\omega}) = \beta zx_z^1 + \beta z x_z^B,
\end{equation}
where we have used the first moment property of the multinomial distribution. Similar expressions are found for the transitions $S \rightarrow I_2$, $I_1 \rightarrow B$, $I_2 \rightarrow B$. Finally, inserting each of the appropriate transition rates into Eq.~\eqref{eq:def:MF_evolution_node} gives the system of mean-field evolution equations:
\begin{align}
    \frac{dx_z^S}{dt} &= -\beta z(x_z^1+x_z^B)x_z^S -\beta z(x_z^2+x_z^B)x_z^S + x_z^1 + x_z^2 \label{eq:MF_CO_SIS_1}\\
    \frac{dx_z^1}{dt} &= -x_z^1 -\lambda\beta z(x_z^2+x_z^B)x_z^1 + \beta z(x_z^1+x_z^B)x_z^S +x_z^B \label{eq:MF_CO_SIS_2}\\
    \frac{dx_z^2}{dt} &= -x_z^2 -\lambda\beta z(x_z^1+x_z^B)x_z^2 + \beta z(x_z^2+x_z^B)x_z^S +x_z^B \label{eq:MF_CO_SIS_3}\\
    \frac{dx_z^B}{dt} &= -2x_z^B +\lambda\beta z(x_z^2+x_z^B)x_z^1 + \lambda\beta z(x_z^1+x_z^B)x_z^2 \label{eq:MF_CO_SIS_4}.
\end{align}
Equations~\eqref{eq:MF_CO_SIS_1} to \eqref{eq:MF_CO_SIS_4} form a closed system that describes the time evolution from initial conditions $x_z^S(0), x_z^1(0), x_z^2(0), x_z^B(0)$.

\begin{figure}[t!]
    \centering
    \includegraphics[width=0.4\textwidth]{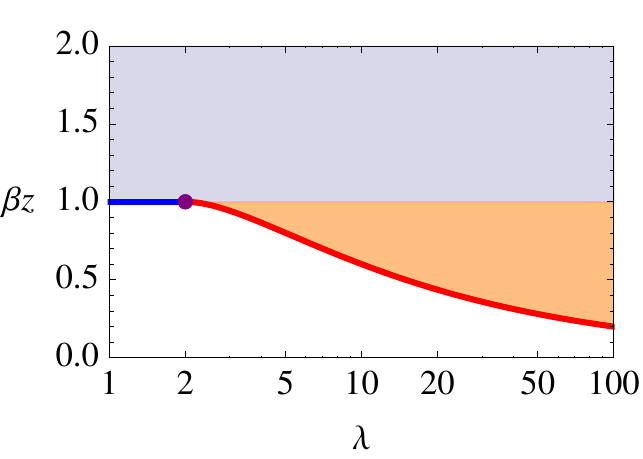}
    \caption{The phase diagram of the cooperative SIS model on a degree-regular network where each node has degree $z$. In all areas of the parameter space, the disease-free state exists.  In the blue area above the line $\beta = 1/z$, the endemic state also exists and is the only stable equilibrium. In the orange region between the line $\beta = 1/z$ and the critical curve of Eq.~\eqref{eq:beta_c_lambda} (thick red curve) both the endemic and disease-free states are stable, with the equilibrium state in this region depending on the initial conditions.}
    \label{fig:Co_SIS_PhasePlane}
\end{figure}

Of particular interest in epidemic dynamics is whether or not the disease eventually dies out, i.e., if the population eventually become disease-free or if it remains in an equilibrium endemic state. The equilibrium behavior of the dynamics can be analyzed by setting the time derivatives of Eqs.~\eqref{eq:MF_CO_SIS_1} to \eqref{eq:MF_CO_SIS_4} to zero and solving the resulting equations for 
$\bar{x}_z^S, \bar{x}_z^1, \bar{x}_z^2$, and $\bar{x}_z^B $, where the bar  denotes steady-state values. The equilibrium states are summarized in the phase diagram of Fig.~\ref{fig:Co_SIS_PhasePlane}, which shows the regions of $(\lambda, \beta)$ space where qualitatively different equilibrium behaviour occurs. We now describe these equilibrium states, and the manner in which they are calculated.

The equilibrium behaviour of the dynamics is analyzed in terms of the total fraction of infected nodes $\bar{i}$ in the network, where $\bar{i} = \bar{x}_z^1+\bar{x}_z^2+\bar{x}_z^B = 1-\bar{x}_z^S$. If $\bar{i}=0$, both diseases eventually dies out; otherwise, one or both diseases  become endemic. The disease-free state is an equilibrium state that exists for all values of $\beta$ and $\lambda$. However, depending on the values of $\beta$ and $\lambda$, this equilibrium state may be either stable or unstable. In the case of an unstable disease-free state, any perturbation (i.e., any non-zero value of the initial infected fractions) will eventually lead to an endemic equilibrium state. 
The stability of the disease-free state is examined through the eigenvalues of the Jacobian matrix associated with the system of equations~\eqref{eq:MF_CO_SIS_1} to \eqref{eq:MF_CO_SIS_4}, evaluated at $x_z^S=1$ and $x_z^1=x_z^2=x_z^B=0$. The largest non-zero eigenvalue\footnote{Note that the zero eigenvalue always exists; however this is an artificial eigenvalue arising from the fact that the system given by Eqs.~\eqref{eq:MF_CO_SIS_1} to \eqref{eq:MF_CO_SIS_4} along with the equation $\bar{x}_z^S+\bar{x}_z^1+\bar{x}_z^2+\bar{x}_z^B=1$ is overspecified.}, $e_{\text{max}}$, is given by
\begin{equation}
    e_{\text{max}} = z\beta - 1
\end{equation}
and so the disease-free state is stable only if $\beta<1/z$ which, we note, is independent of the accentuation parameter $\lambda$. Thus, for $\beta<1/z$, the system will converge to the disease-free state when the initial conditions are sufficiently close to this state. 

\begin{figure}
    \centering
    \includegraphics[height = 4cm]{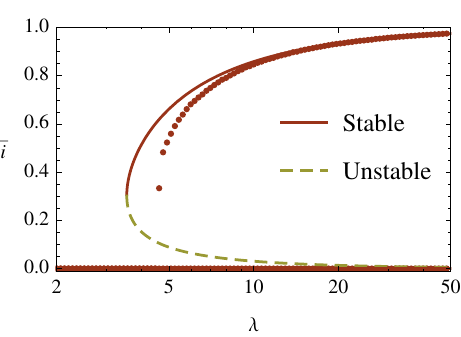} \hspace{8pt}
    \includegraphics[height = 4cm]{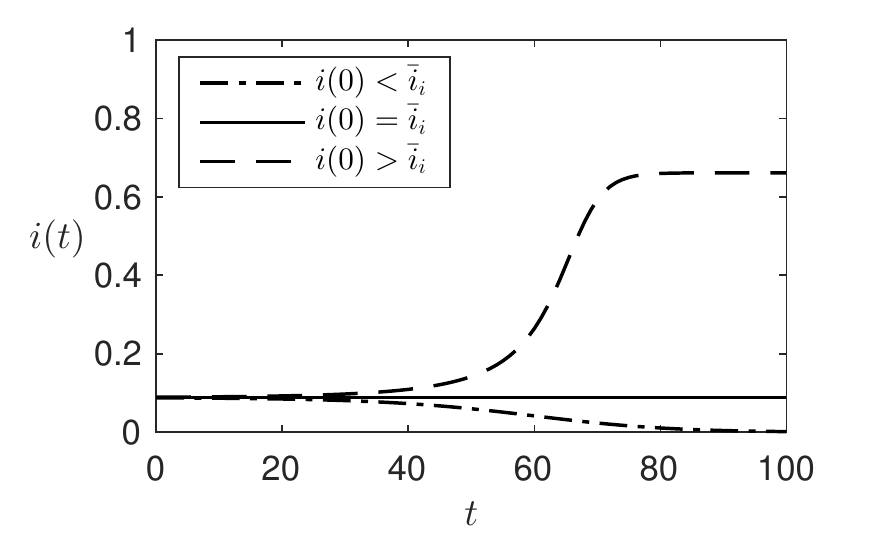}
    \caption{Behaviour of the co-operative SIS model when the disease-free state is stable, i.e., $\beta < 1/z$. Here, the network is a degree-regular network where each node has degree $z=8$, while $\beta$ is fixed to the value $\beta=0.9/z$. Left: The MF theory (curves) predicts the existence of a discontinuous transition when the accentuation parameter $\lambda$ is increased above the critical point $\lambda_c(\beta)$, where $\lambda_c(\beta)$ is calculated by inverting Eq.~\eqref{eq:beta_c_lambda}. Above this critical point there are three possible steady-states -- the disease-free stable state $\bar{i}=0$, the unstable endemic state $\bar{i}=\bar{i}_{-}$ and the stable endemic state $\bar{i}=\bar{i}_{+}$. The qualitative accuracy of the MF predictions are confirmed here by numerical simulations (points). Right: Evolution of the dynamics in the bistable regime $\lambda > \lambda_c(\beta)$ from initial conditions less-than (dot-dashed curve), equal-to (solid curve) and greater-than (dashed curve) the unstable endemic state $\bar{i}=\bar{i}_{-}$, illustrating the reliance of the dynamics on initial conditions.}
    \label{fig:Co_SIS_DiscTran}
\end{figure}

However, other stable endemic states can exist alongside the stable disease-free state in this area of the parameter space $\beta<1/z$, and so an endemic state may be reached depending on the value of the initial conditions. To find the other equilibrium states, we solve the steady-state equations of Eqs.~\eqref{eq:MF_CO_SIS_1} to \eqref{eq:MF_CO_SIS_4}. These equations yield $\bar{x}_z^B=(1-\bar{x}_z^S)(1-z\beta \bar{x}_z^S)/(1+z\beta \bar{x}_z^S)$, $\bar{x}_z^1 = (1-z\beta \bar{x}_z^S)/z\beta\lambda$ and $\bar{x}_z^2=\bar{x}_z^1$, and  along with the condition that $\bar{x}_z^S+\bar{x}_z^1+\bar{x}_z^2+\bar{x}_z^B=1$, this results in two possible values for $\bar{i}$, given by
\begin{align}
    \bar{i}_{+} &= \frac{\lambda-2}{2(\lambda-1)} + \frac{\sqrt{(z\lambda\beta)^2-4(\lambda-1)}}{2z\beta(\lambda-1)}, \\
    \bar{i}_{-} &= \frac{\lambda-2}{2(\lambda-1)} - \frac{\sqrt{(z\lambda\beta)^2-4(\lambda-1)}}{2z\beta(\lambda-1)}.
    \label{eq:endemic_states}
\end{align}
Physically relevant values of $\bar{i}$ must be real and lie between 0 and 1; 
such solutions for $\bar{i}_{+}$ and $\bar{i}_{-}$ exist only if $(z\lambda\beta)^2-4(\lambda-1)\geq0$ and $\lambda \geq 2$. The epidemic threshold is therefore the smallest value of $\beta$ for which these constraints are satisfied, and so the epidemic threshold $\beta_c$ as a function of the accentuation parameter $\lambda$ is given by
\begin{equation}
    \;\;\;\;\;\; \beta_c(\lambda) = \frac{2\sqrt{\lambda-1}}{z\lambda}, \;\;\;\;\;\; \lambda\geq 2.
    \label{eq:beta_c_lambda}
\end{equation}
This $\beta_c(\lambda)$ epidemic threshold curve is illustrated by the thick red curve in the phase diagram of Fig.~\ref{fig:Co_SIS_PhasePlane}. Note that the value of $\beta_c(\lambda)$ is always less than $1/z$, which is the epidemic threshold in the absence of interaction from Eq.~\eqref{eq:ET1disease}. It follows that interaction between the diseases lowers the value of the epidemic threshold, allowing endemic states to exist even when the infection transmission rates $\beta$ are relatively small. Strikingly, the transition across this threshold is discontinuous, in contrast to the continuous transition that occurs in traditional SIS dynamics. At the point $\beta_c(\lambda)$ the fraction of infected nodes $\bar{i}$ can jump from zero to a finite value $(\lambda-2)/2(\lambda-1)$. This is illustrated in the left panel of Fig.~\ref{fig:Co_SIS_DiscTran}, which also shows how small changes in the value of $\lambda$ --- when close to the critical value $\lambda_c(\beta) = 2\left( 1+ \sqrt{1-(z\beta)^2}\right)/(z\beta)^2$ obtained from inverting Eq.~\eqref{eq:beta_c_lambda} --- can trigger large changes in the macroscopic behaviour of the system.

Previously, we saw that the disease-free state was unstable when $\beta > 1/z$. In this region of the parameter space, only one of the two solutions given by Eq.~\eqref{eq:endemic_states}, the $\bar{i}_{+}$ solution, is physically meaningful. This is the stable solution, and so for any initial condition satisfying $x_z^S(0)<1$ the equilibrium fraction of infected nodes will be given by $\bar{i}_{+}$. In the region of parameter space given by $\beta_c(\lambda) < \beta < 1/z$, both the $\bar{i}_{+}$ and $\bar{i}_{-}$ solutions exist and there are two stable states, the disease-free state and the endemic state $\bar{i}_{+}$, along with the unstable state $\bar{i}_{-}$. Here, the initial conditions of the system affect the steady-state, as initial values $i(0) < \bar{i}_{-}$ will tend to the disease-free state and values $i(0) > \bar{i}_{-}$ will tend to the endemic state. The right panel of Figure~\ref{fig:Co_SIS_DiscTran} illustrates this behaviour, showing how the dynamics evolve to the three different steady states from three different initial conditions, $i(0)>\bar{i}_{-}$, $i(0) = \bar{i}_{-}$ and $i(0)<\bar{i}_{-}$. It is also noteworthy that the unstable steady state $\bar{i}_{-}$ tends monotonically to zero as a function of $\lambda$ as $\mathcal{O}(\lambda^{-1})$ for large  $\lambda$ (as seen from Eq.~\eqref{eq:endemic_states}), indicating that for large values of $\lambda$ the disease-free state will only be reached if the initial seed fraction of infected nodes is negligibly small.

The multistate MF framework of Eq.~\eqref{eq:def:MF_evolution_node} has given us key insights into the behaviour of co-operative SIS dynamics on degree-regular networks. The most notable property of the dynamics revealed by our analysis is the existence of a bistable equilibrium regime where both a disease-free stable state and endemic stable state exist, and the discontinuous transition that can occur upon entering this area of the parameter space. The simplicity of the model and of the MF framework have been beneficial for several reasons. Firstly, certain analyses of more complicated, multi-parameter models of co-operative SIS processes --- of which this model is a special case --- have not revealed the existence of discontinuous-transition behaviour, as the large dimensionality of the system can be restrictive in finding this type of behaviour~\cite{sanz2014dynamics}. Secondly, cases where discontinuous behaviour has been identified have been largely restricted to study by numerical simulations~\cite{hebert2015complex}, and the understanding that this gives regarding the behaviour of the system in the bistable regime is limited. The MF analysis not only reveals the discontinuous transitions associated with co-operative SIS dynamics --- and the approximate locations in parameter space at which these transitions occur --- but it also gives analytical expressions for both the stable and unstable equilibrium states in the bistable equilibrium regime, therefore increasing understanding of the behaviour in the bistable regime. 

The knowledge of the unstable state is of particular importance, as it dictates the final equilibrium state that the system will reach and gives a measure of the size of the perturbation required to shift the system from the disease-free state to the endemic state. This perturbation must be quite large for values of $\lambda$ close to the critical point, and so the system should robustly stay in the disease-free state for such relatively small values of $\lambda$. However, the size of the required perturbation decreases to zero as $\lambda$ increases, and so for sufficiently large values of $\lambda$ only small perturbations from the disease-free state are required to move the system into the state of a large endemic outbreak. Our analysis has therefore given very detailed insights into the potential for catastrophic-type behaviour in co-operative SIS dynamics, and such behaviour has implications for the prediction and control of real diseases that interact in a co-operative manner~\cite{fleming1999epidemiological, singer2009introduction}.
 
To complete our study of the co-operative SIS model, we examine the effect of heterogeneity in the degree distribution on the dynamics of the system. In Appendix~\ref{sec:ET_CoSIS_hetero}, we show that degree heterogeneity \emph{promotes} the existence of an endemic state. There, we derive the critical point 
\begin{equation}
    \beta_c = \frac{z}{\langle k^2 \rangle}
    \label{eq:ETthredhold}
\end{equation}
where for values $\beta > \beta_c$ the endemic state will always exist. This condition of Eq.~\eqref{eq:ETthredhold} implies that $\beta_c \rightarrow 0$ for networks with degree distributions such that $\langle k^2 \rangle \rightarrow \infty$ (such as power-law distributions $p_k \propto k^{-\gamma} $ for $ 2 < \gamma < 3 $). In such cases, the endemic state will exist for any value $\beta>0$ and thus these networks are highly susceptible to endemic outbreaks. Interestingly, the threshold condition of Eq.~\eqref{eq:ETthredhold} does not depend on $\lambda$, and so in networks with $\langle k^2 \rangle \rightarrow \infty$ even small values $\lambda \ll 1$ (in which case infection with one disease implies relative immunity towards the other) will not constrain an outbreak.

\subsection{Kinetically constrained dynamics and the necessity of high accuracy frameworks}
\label{chpt6:sec3:relevance}

In Section~\ref{sec:CoSIS} the multistate MF equations accurately captured the dynamics of the cooperative disease model of Eq.~\eqref{eq:CoSISratemat}, providing a basis for  insightful analysis into the behaviour of the model. However, MF approximation frameworks will not always provide a sufficient level of accuracy, as the assumption of dynamical independence between each node in the network may be too strong. Higher accuracy frameworks such as the PA and AME are required in such cases to provide tools for analysis, and we now illustrate this point with the study of a complex spin glass model from the physical sciences domain.

The Fredrickson-Andersen (FA) model is a spin model of glass-forming liquids~\cite{fredrickson1984kinetic}, liquids that when supercooled from high temperatures can form crystalline structures~\cite{biroli2012random}. In the FA model, a lattice or network represents the physical substrate, with nodes in the network having a spin which is either positive (+1) or negative (-1) representing dense and sparse areas of the substrate respectively. A node's state can dynamically change over time but does so according to a kinetic constraint: state changes can only occur if the number of neighbours of a node with negative spin is greater than or equal to $f$, where $f$ is the so-called facilitation parameter of the system. This kinetic constraint mechanism mimics jamming, where nodes can only be active (mobile) if they have space to do so because of a sufficiently number of spin -1 ("sparse") neighbours. Once this constraint is met, nodes with spin -1 will change to spin +1 at a rate 1, while nodes with spin +1 will change to spin -1 at a rate $e^{-1/T}$; the latter rate is less than one and so the system favours the existence of spin +1 nodes. Combining the kinetic constraint with the spin flip rates gives the state transition rates of the system:
\begin{align}
    &F_{\mathbf{m}}(-1 \rightarrow +1) = H(m_{-1} - f) \\
    &F_{\mathbf{m}}(+1 \rightarrow -1) = H(m_{-1} - f)\times e^{-1/T},
\end{align}
where $H(x)$ is the Heavyside step function that takes the value $H(x)=1$ for non-negative values of $x$ and zero otherwise.

\begin{figure}[t!]
    \centering
    \includegraphics[height = 4cm]{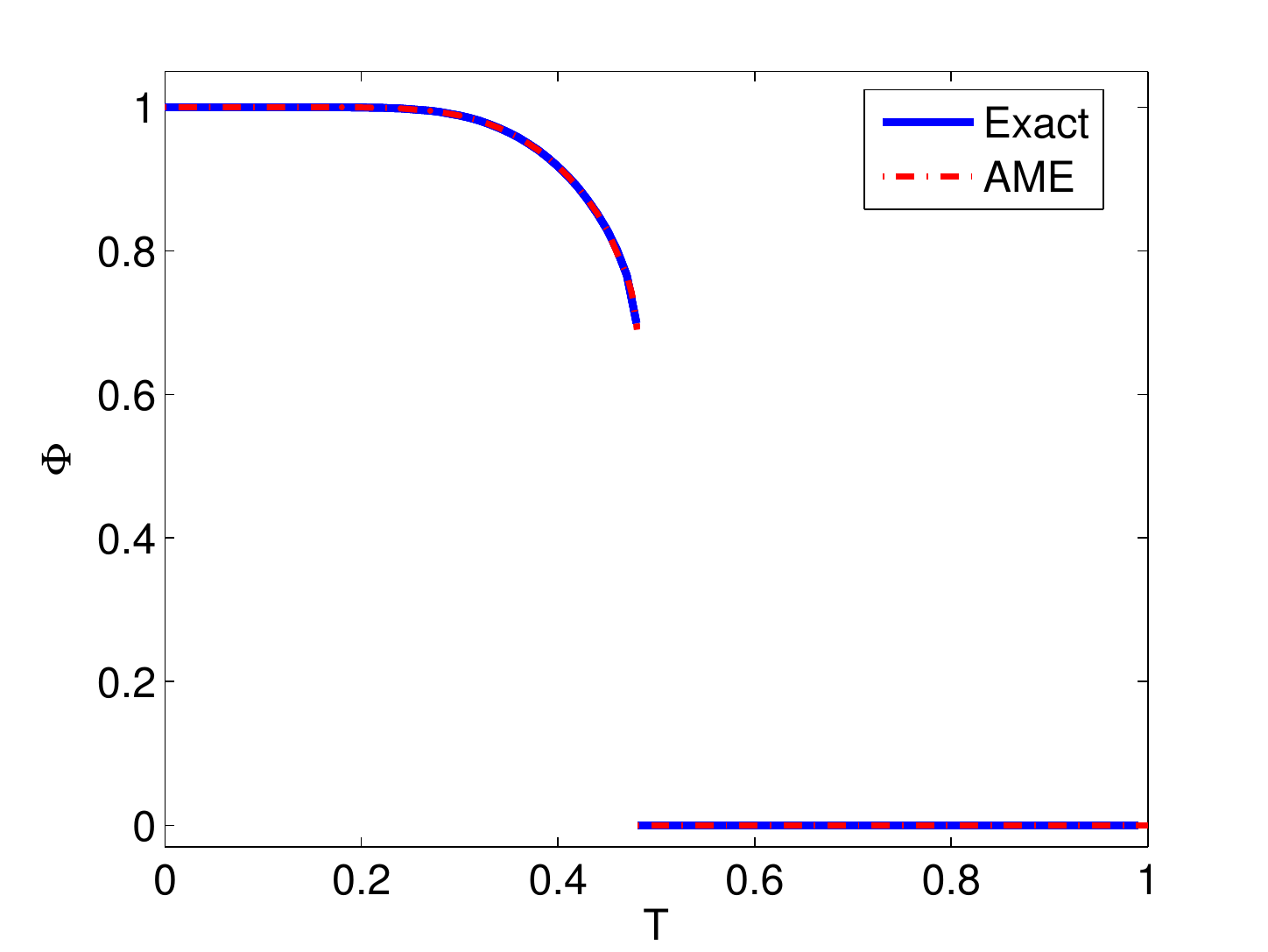} \hspace{10pt}
    \includegraphics[height = 4cm]{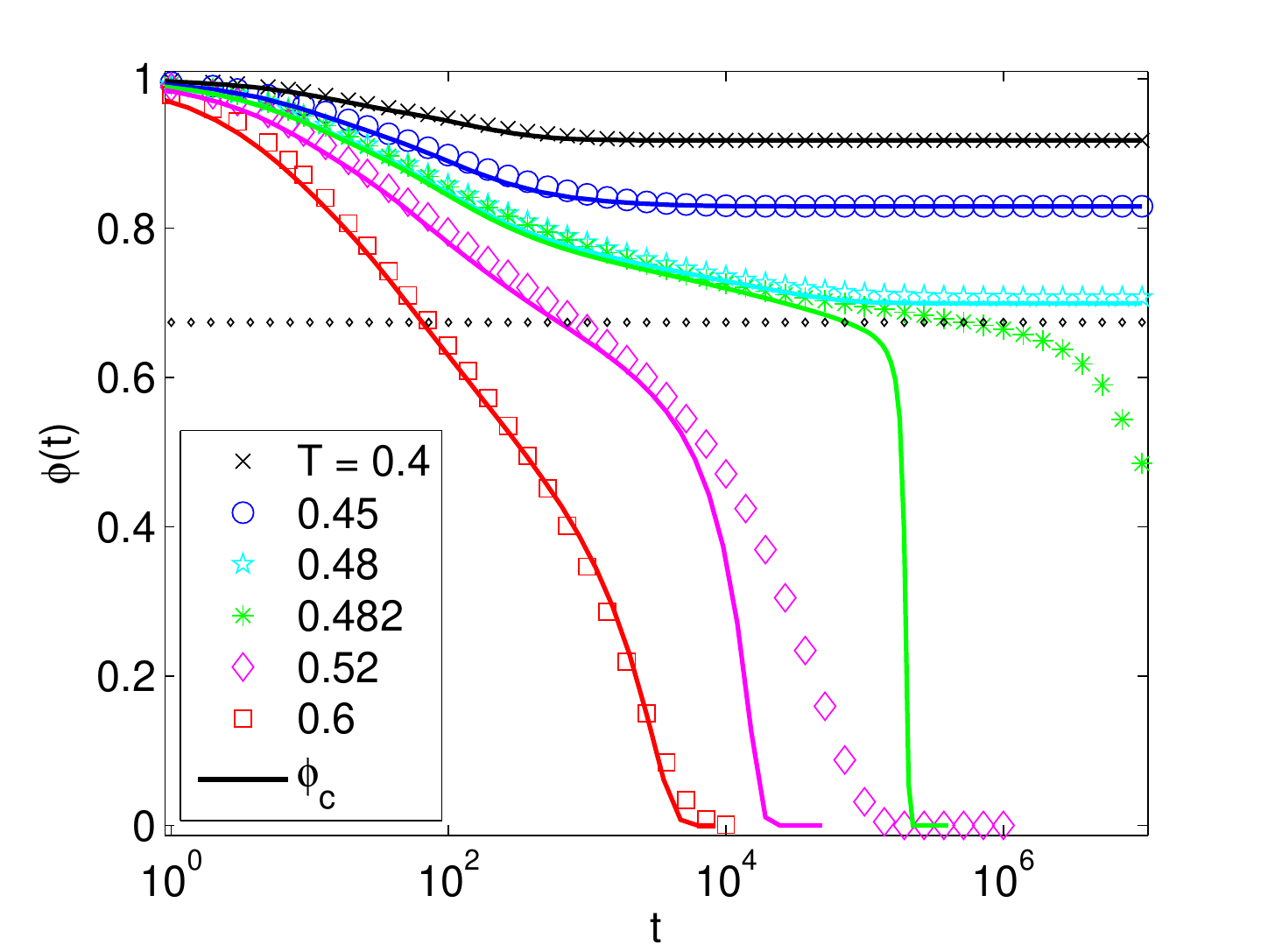}
    \caption{Behaviour of the FA model on degree-regular networks ($z=4$) with facilitation parameter $f=2$ for various values of the temperature parameter $T$. Left: A discontinuous transition of the equilibrium order parameter occurs  as the temperature $T$ is reduced below the critical temperature $T_c$. The exact solution here is the equilibrium branching process approach on tree-like networks presented by Sellito \emph{et al.}~\cite{sellitto2010}. Right: The four-state AME (solid lines) captures the transient behaviour of the FA model (symbols show  numerical simulation results), recreating both the glassy and liquid states. The transient variable $\phi(t)$ here is the fraction of frozen spins in the network at time $t$, with $\lim_{t\rightarrow \infty}\phi(t) = \Phi$}.
    \label{fig:FAmodel}
\end{figure}

The FA model is one of a class of facilitated spin models, so called because mobile nodes (i.e., nodes that are not jammed) can facilitate the mobilization of other nodes that are jammed as a result of the kinetic constraint. In this fashion, facilitation can recursively cause the mobilization of jammed nodes and the relaxation of system, in which case the resulting equilibrium state is the liquid state. On the other hand, the system may not fully relax, with jammed nodes occupying much of the substrate. This is the glass state, where the system is in dynamical arrest consisting of both mobile and frozen nodes. The final equilibrium state depends heavily on the temperature $T$, with a critical temperature $T_c$ separating the liquid phase ($T>T_c$) and the arrested glassy phase ($T_c < T$) (Fig.~\ref{fig:FAmodel}). The essential quantity or order parameter that defines such states is the fraction $\Phi$  of frozen nodes in the system at equilibrium, with $\Phi=0$ in the liquid state and $\Phi > 0$ in the glassy state.

In~\cite{fennell2014analytical}, we showed that despite the fact that the FA dynamics have nodes with only two possible spins $+1$ or $-1$, binary-state theoretical frameworks, including the high-accuracy 2-state AME of Refs.~\cite{gleeson2011high,gleeson2013},  incorrectly predict the equilibrium value of $\Phi$ to be zero for all values of the temperature, thus failing to capture the existence of the glassy phase at low temperatures.
We therefore extended the state space of the dynamics to four states by including an auxiliary state $c$ or $u$ for each node which defines whether the node's spin state had changed since $t=0$  ($c$) or was as yet unchanged ($u$). The corresponding rate matrix function for the 4-state dynamics is
\begin{equation}
  \mathbf{F}_{\mathbf{m}}=
  \bordermatrix{
    \hspace{4pt} & (-1,u), & (+1,u) & (-1,c) & (+1,c)  \vspace{4pt} \cr
    \vspace{4pt}
    (-1,u) \hspace{4pt} & 0 & 0 & 0 & 1 & \cr
    \vspace{4pt}
    (+1,u) \hspace{4pt} & 0 & 0 & e^{-1/T} & 0 \cr
    \vspace{4pt}
    (-1,c) \hspace{4pt} & 0 & 0 & 0 & 1  \cr
    \vspace{4pt}
    (+1,c) \hspace{4pt} & 0 & 0 & e^{-1/T} & 0 \cr
  }*H(m_{(-1,u)}+m_{(-1,c)}-f)
  \label{eq:FAratemat}
\end{equation}
By solving the system of equations given by Eq.~\eqref{eq:def:master_equation_multistate}, we found that the 4-state AME can capture the essential dynamics of the system to a high level of accuracy. The 4-state AME not only qualitatively predicts the existence of two equilibrium phases corresponding to liquid and glass, but also provides  accurate predictions of  the critical temperature $T_c$ that separates the two phases and the values of the order parameter $\Phi$ in the glassy phase (Fig~\ref{fig:FAmodel}, left panel). Furthermore, the 4-state AME  accurately reproduces the transient dynamics of the system (Fig~\ref{fig:FAmodel}, right panel), a challenging task due to the long range correlations between distant nodes at temperatures close to the critical point $T_c$. The 4-state AME captures the spatial heterogeneity of the dynamics in the glass phase, including the clustering of mobile and blocked nodes, which results in ``patchy" equilibrium configurations.

\begin{figure}[t!]
    \centering
    \includegraphics[height = 4cm]{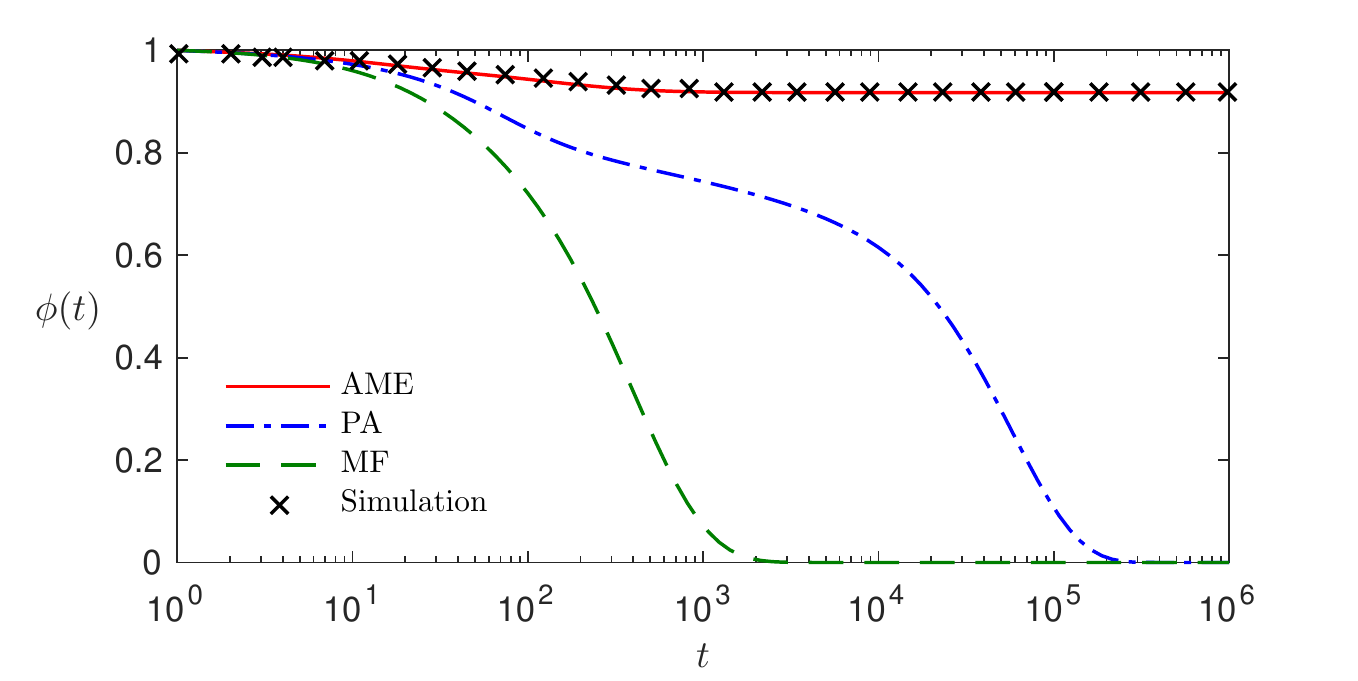} 
    \caption{Comparison of the AME, PA and MF frameworks for transient FA dynamics. As in Fig.~\ref{fig:FAmodel}, the network is degree-regular ($z=4$) with facilitation parameter $f=2$. The temperature here is chosen to be below the critical point ($T=0.4$). The AME matches excellently to the numerical simulation results, while the MF and PA fail to capture the model behaviour. These qualitative observations hold for all values $T < T_c$.}
    \label{fig:PA_MF_fail}
\end{figure}

On the contrary, the rich behaviour of the FA model cannot be captured by the 4-state PA and MF frameworks. Like their binary-state counterparts, these frameworks fail to reproduce the frozen glassy state, predicting that the fraction of frozen nodes $\Phi$ at equilibrium is zero for all values of the temperature $T$. We illustrate this by examining the system at a temperature $T<T_c$ as shown in Fig.~\ref{fig:PA_MF_fail}. Here, the fraction of frozen nodes in the system as predicted by the AME converges to the equilibrium value $\Phi \approx 0.91$, a value matched closely by numerical simulation of the dynamics. This value is positive, indicating that the system is in the glassy phase. However, the fraction of frozen nodes as predicted by both the PA and MF frameworks converges to $\Phi=0$, thus predicting that the system is in  the liquid phase. Examination of the 4-state MF and PA equations indicates that $\Phi>0$ is not a possible equilibrium for any value of the temperature $T$, and thus they are insufficient for the study of these rich dynamics. 

So why does the 4-state AME capture the dynamics of the FA model while the 4-state PA and MF do not? The answer lies in the non-linear relationship between the states of nodes and the states of their neighbours. The AME variables $x^i_{\mathbf{m}}$ capture the relationship between a node and all of its neighbours, while the PA and MF frameworks approximate this relationship by link relationships and the multinomial distribution as given by Eqs.~\eqref{eq:def:PA_ansatz}, \eqref{eq:def:Mult} and \eqref{eq:def:Mult_2}. Analysis of the AME equations for the FA model reveal non-linear relationships; specifically the variables $x^{(-1,u)}_{\mathbf{m}}$ and $x^{(+1,u)}_{\mathbf{m}}$ satisfy
\begin{align}
    x^{(-1,u)}_{\mathbf{m}}, x^{(+1,u)}_{\mathbf{m}} \mbox{ are }
    \begin{cases}
        = 0 & \text{ if } m_{(-1,u)} + m_{(-1,c)} \geq f, \\
        > 0 & \text{ if }m_{(-1,u)} + m_{(-1,c)} < f. \\
    \end{cases}
\end{align}
This non-linear relationship cannot be captured by the multinomial distribution approximation, whose values vary smoothly as a function of $m_{(-1,u)} + m_{(-1,c)}$. Indeed, the non-linearity is essential for the freezing mechanism. The variables $x^{(-1,u)}_{\mathbf{m}}$ and $x^{(+1,u)}_{\mathbf{m}}$ cannot be non-zero for $m_{(-1,u)} + m_{(-1,c)} \geq f$, (as otherwise they are mobile and will change state from $u$ to $c$), while they must be non-zero for $m_{(-1,u)} + m_{(-1,c)} < f$ to allow for the existence of a non-zero fraction of frozen nodes in the glassy regime. The 4-state PA and MF do not account for this non-linearity, and as a result  their evolution equations for $x_k^{(-1,u)}$ and $x_k^{(+1,u)}$ do not reach a steady state until $x_k^{(-1,u)} = x_k^{(+1,u)}$, thus eliminating the possibility of a glassy equilibrium state. 

Interestingly, our analysis here can be related to earlier results presented by Gleeson in~\cite{gleeson2013} when examining binary-state AME, PA and MF frameworks. There, the equilibrium states of the PA and AME frameworks were equivalent for a spin model known as the Ising model, a model that can recover the FA model in the absence of kinetic constraint (i.e., $f \rightarrow \infty)$. However, the AME was the only framework capable of capturing the dynamics of threshold models which, as described in Section~\ref{sec:multistate_dynamics}, have similar constraints to the FA model on state changing. The introduction of the kinetic constraint in the FA model has a large effect on the complexity of the dynamics and thus in the ability of various approximations frameworks to adequately capture its dynamics. Our observations here have important implications for the study of models with kinetic constraints. Threshold models are widely employed models of complex contagion in the social sciences~\cite{granovetter1973strength,centola2007complex,morone2015influence,iniguez2017service}, and the AME framework that we have introduced here lends itself to the analysis of the traditional binary-state threshold models~\cite{watts2002simple} as well of more complex multistate threshold models that are the focus of much current attention~\cite{melnik2013, kuhlman2015limit}.

\section{Conclusions}
\label{sec:conclusions}
In this paper we have derived degree-based approximation frameworks for the analysis of a wide class of multistate Markovian dynamical processes that can be defined in terms of rate matrix functions $\mathbf{F}_{\mathbf{m}}$. These frameworks provide a  comprehensive set of tools for the analysis of multistate dynamical processes on networks, and can be used to study specific dynamics  through an appropriate expression of a rate matrix function.

The multistate frameworks we introduced are of varying levels of accuracy, and this provides flexibility in the choice of framework that should be employed when studying a dynamical process. High accuracy frameworks can be used to study complex processes whose dynamics cannot be captured by lower accuracy methods, a feature that we illustrated through our analysis of the Fredrickson-Andersen model of glassy dynamics in Section~\ref{chpt6:sec3:relevance}. On the other hand, lower accuracy frameworks such as mean-field, when capable of capturing the essential aspects of the dynamics, can be sufficiently tractable to allow analytical insight. The power of such analytically tractable frameworks has been illustrated in our study of the cooperative disease model in Section~\ref{sec:CoSIS}, as the mean-field framework gives understanding of the combined effect of the dynamical parameters and the network connectivity on the equilibrium behaviour of the dynamics.

We believe that our multistate frameworks provide a useful contribution to the field of dynamical processes on networks, and thus to the understanding of real dynamics in a wide range of applications. Furthermore, we have made optimized code available to produce numerical solutions to the equations of the approximation frameworks~\cite{website:MultiAMEcode}, giving our work an extra level of accessibility to the community at large.

\section*{Acknowledgments}
This work has been supported by Science Foundation Ireland, Grants No. 11/PI/1026 and No. 16/IA/4470, and by the James S. McDonnell Foundation. We acknowledge helpful discussions with Fakhteh Ghanbarnejad and Davide Cellai.

\appendix

\section{Calculation of the cooperative SIS epidemic threshold on networks with heterogeneous degree distributions}
\label{sec:ET_CoSIS_hetero}

In this Appendix, we examine the behaviour of the co-operative SIS model as defined in Section~\ref{sec:CoSIS} on networks with heterogeneous degree distributions $p_k$.
The mean-field equations for this system are obtained by inserting the rate function of Eq.~\eqref{eq:CoSISratemat} into Eqs.~\eqref{eq:def:MF_evolution_node}; this gives the set of equations
\begin{align}
    \frac{dx^S_k}{dt} &= -\beta k(w^{I_1}+w^B)x^S_k -\beta k(w^{I_2} +w^B)x^S_k + x^{I_1}_k + x^{I_2}_k, \label{eq:MF_CO_X^0_KIX^0_K_1}\\
    \frac{dx^{I_1}_k}{dt} &= -x^{I_1}_k -\lambda\beta k(w^{I_2}+w^B)x^{I_1}_k + \beta k(w^{I_1}+w^B)x^S_k +x^B_k ,\label{eq:MF_CO_X^0_KIX^0_K_2}\\
    \frac{dx^{I_2}_k}{dt} &= -x^{I_2}_k -\lambda\beta k(w^{I_1}+w^B)x^{I_2}_k + \beta k(w^{I_2}+w^B)x^S_k +x^B_k, \label{eq:MF_CO_X^0_KIX^0_K_3}\\
    \frac{dx^B_k}{dt} &= -2x^B_k +\lambda\beta k(w^{I_2}+w^B)x^{I_1}_k
                        + \lambda\beta
                        k(w^{I_1}+w^B)x^{I_2}_k. \label{eq:MF_CO_X^0_KIX^0_K_4}
\end{align}
which holds for all values $k_{\min}\leq k \leq k_{\max}$. To analyze the system, we employ the change of variables 
\begin{align}
	x^S_k &\leftarrow x^S_k \\
	y_k &\leftarrow x^{I_2}_k+ x^{I_1}_k \\
	\Delta_k &\leftarrow x^{I_2}_k-x^{I_1}_k \\
	x^B_k &\leftarrow x^B_k,
\end{align}
in which case we arrive at the set of equations
\begin{align}
    \frac{dx^S_k}{dt} &= -\beta k(w^y+2w^B)x^S_k + y_k \\
    \frac{dy_k}{dt} &=  -y_k -\lambda\beta k \left(\frac{w^yy_k - w^{\Delta}\Delta_k}{2} + w^By_k\right) + \beta k(w^y+2w^B)x^S_k + 2x_k^B\\
    \frac{d\Delta_k}{dt} &= -\Delta_k -\lambda\beta k \left(\frac{w^y\Delta_k - w^{\Delta}y_k}{2} + w^B \Delta_k \right) + \beta kw^{\Delta}x^S_k \\
    \frac{dx^B_k}{dt} &= -2x^B_k + \lambda\beta k \left(\frac{w^yy_k - w^{\Delta}\Delta_k}{2} + w^By_k\right).
\end{align}
Using the property $x^B_k=1-x^S_k-y_k$ (and thus $w^B=1-w^S-w^y$), and restricting the equations to the steady state we have the following system of equations which allow for the analysis of the equilibrium behaviour:
\begin{align}
    0 &= -\beta k(2-2w^S-w^y)x^S_k + y_k \label{eq:APP1} \\
    0 &=  -y_k - \lambda\beta k \left((1-w^S)y_k - \frac{w^yy_k +
                      w^{\Delta}\Delta_k}{2} \right) \label{eq:APP2}
  \\ &+ \beta k(2-2w^S-w^y)x^S_k + 2(1-x^S_k-y_k) \nonumber \\
    0 &= -\Delta_k- \lambda\beta k \left( (1-w^S) \Delta_k - \frac{w^y\Delta_k + w^{\Delta}y_k}{2} \right) + \beta kw^{\Delta}x^S_k \label{eq:APP3}.
\end{align}
From Eq.~\eqref{eq:APP1} we obtain expressions for $y_k$ and $w^y$ in terms of the other variables as
\begin{align}
    y_k &= 2\frac{\beta k(1-w^S)}{1+\beta U_S}x^S_k  \\
    w^y &= 2\frac{(1-w^S) \beta U_S}{1+\beta U_S},
\end{align}
where $U_S = \sum_k (kp_k/z)x_k^S$, while Eq.~\eqref{eq:APP3} lets us isolate $\Delta_k$ as
\begin{align}
  \Delta_k  =\frac{\beta k w^{\Delta} \left(1+\lambda\beta
              k\frac{1-w^S}{1+\beta U_s}\right)}{1+\lambda\beta k \frac{
              1-w^S}{1+\beta U_s}}x^S_k = \beta k w^{\Delta}x^S_k.
\end{align}
Finally, $y_k$, $w^y$ and $\Delta_k$ can be inserted into Eq.~\eqref{eq:APP2} to give
\begin{align}
0 &= \left(-\left(\lambda\beta k\frac{1-w^S}{1+\beta
U_s}+2\right)\left( \frac{\beta k(1-w^S)}{1+\beta U_s}\right)+ \left(\left(\lambda\beta k\frac{w^{\Delta}}{4}\right) \beta k w^{\Delta}\right) -1\right)x^S_k + 1,
\end{align}
and thus we have that
\begin{align}
 x^S_k &= \frac{1}{ 1+ 2\frac{\beta k(1-w^S)}{1+\beta U_s} + \lambda(\beta k)^2\left(\left(\frac{1-w^S}{1+\beta
U_s}\right)^2 - \left(\frac{w^{\Delta}}{2}\right)^2\right)}
\end{align}
and
\begin{align}
 w^S &=\sum_k\frac{kp_k}{z}\frac{1}{ 1+ 2\frac{\beta k(1-w^S)}{1+\beta U_s} + \lambda(\beta k)^2\left(\left(\frac{1-w^S}{1+\beta
U_s}\right)^2 - \left(\frac{w^{\Delta}}{2}\right)^2\right)}.
\label{eq:APP4}
\end{align}

We can see that $w^S=1$ (in which case $w^{\Delta}=0$) is a solution to Eq.~\eqref{eq:APP4}. This is the trivial solution that always exists in which case every node in the network is healthy. We search for an endemic solution $0 \leq w^S < 1$ by subtracting 1 from both sides of Eq.~\eqref{eq:APP4} and dividing by $w^S-1$, giving
\begin{align}
 1&=\sum_k\frac{kp_k}{z}\frac{2\frac{\beta k}{1+\beta U_s} + \lambda(\beta k)^2\left(\frac{1-w^S}{(1+\beta
U_s)^2} - \frac{1}{1-w^S}\left(\frac{w^{\Delta}}{2}\right)^2\right)}{ 1+ 2\frac{\beta k(1-w^S)}{1+\beta U_s} + \lambda(\beta k)^2\left(\left(\frac{1-w^S}{1+\beta
U_s}\right)^2 - \left(\frac{w^{\Delta}}{2}\right)^2\right)} 
\label{eq:APP5}
\end{align}
Now, to see if there is a solution $0\leq w^S<1$ to Eq.~\eqref{eq:APP5}, we examine the right hand side of Eq.~\eqref{eq:APP5} at the limits $w^S=0$ and $w^S \rightarrow 1$ -- a solution is guaranteed to exist of one of these values is less than 1 and one greater than 1 (as Eq.~\eqref{eq:APP5} is a continuous function of $w^S$). The right hand side of Eq.~\eqref{eq:APP5} in the limit $w^S \rightarrow 1$ (and so
$U_s \rightarrow \langle k^2 \rangle/z$ and $w^{\Delta}\rightarrow 0$) is
\begin{align}
\lim_{w^S\rightarrow 1}rhs(w^S)&=\sum_k\frac{kp_k}{z}\left(\frac{2\beta k}{1+\beta
              \langle k^2 \rangle/z} + \frac{\lambda(\beta
              k)^2}{4}\lim_{w^S\rightarrow
              1}\frac{\left(w^{\Delta}\right)^2}{1-w^S}\right).
              \label{eq:rhsw=1}
\end{align}
Now
\begin{equation}
 \left(w^{\Delta}\right)^2 =
 \left(\sum_k\frac{kp_k}{z}(x_k^{I_2}-x_k^{I_1})\right)^2 \leq
 \left(\sum_k\frac{kp_k}{z}\max(x_k^{I_2},x_k^{I_1})\right)^2 
\end{equation}
and 
\begin{equation}
  1-w^S = \sum_k\frac{kp_k}{z}(x_k^{I_1} + x_k^{I_2} +x_k^{B}) \geq
  \sum_k\frac{kp_k}{z}\max(x_k^{I_2},x_k^{I_1}) 
\end{equation}
and so
\begin{align}
  \lim_{w^S\rightarrow 1}\frac{\left(w^{\Delta}\right)^2}{1-w^S} &\leq \lim_{w^S\rightarrow 1}
  \frac{\left(\sum_k\frac{kp_k}{z}\max(x_k^{I_2},x_k^{I_1})\right)^2 }
                                                      {\sum_k\frac{kp_k}{z}\max(x_k^{I_2},x_k^{I_1})} 
  \\ &= \lim_{w^S\rightarrow
  1}\sum_k\frac{kp_k}{z}\max(x_k^{I_2},x_k^{I_1}) = 0,
\end{align}
as $x_k^{I_1}, x_k^{I_2} \rightarrow 0 $ when $w^S \rightarrow 1$. Thus, from Eq.~\eqref{eq:rhsw=1} we have that
\begin{align}
 \lim_{w^S\rightarrow
  1}rhs(w^S)&=\sum_k\frac{kp_k}{z}\frac{2\beta k}{1+\beta
              \langle k^2 \rangle/z} \\ 
  &= \frac{2\beta\langle k^2 \rangle/z}{1+\beta \langle k^2 \rangle/z}. \label{eq:APP6}
\end{align}
On the other hand, in the case that $w^S=0$ (and so
$U_s = 0)$ we have 
\begin{align}
 rhs(w^S=0)&=\sum_k\frac{kp_k}{z}\frac{2\beta k + \lambda(\beta
             k)^2\left(1- \left(\frac{w^{\Delta}}{2}\right)^2\right)}{
             1+ 2\beta k + \lambda(\beta k)^2\left(1 -
             \left(\frac{w^{\Delta}}{2}\right)^2\right)} < 1.
\end{align}
Since $rhs(w^S=0)<1$ then a solution $0\leq w^S<1$  is guaranteed to exist if $rhs(w^S\rightarrow 1)>1$; from Eq.~\eqref{eq:APP6} this occurs when
\begin{equation}
  \beta > \frac{z}{\langle k^2 \rangle}.
  \label{eq:APPthresh}
\end{equation}
Thus we have derived the threshold condition, and so when the infection parameter $\beta$ satisfies Eq.~\eqref{eq:APPthresh} there will always exist an endemic state.

\newpage

\bibliographystyle{siamplain}
\bibliography{library}
\end{document}